\documentclass[12pt,preprint]{aastex}
\usepackage{natbib,makeidx}
\usepackage{lscape}
\usepackage{epsfig}
\usepackage{color}
\usepackage{subfigure}

\newcommand{\Ha}{\mbox{H$\alpha$}}
\newcommand{\Hb}{\mbox{H$\beta$}}
\newcommand{\ha}{\mbox{H$\alpha$}}
\newcommand{\hb}{\mbox{H$\beta$}}
\newcommand{\Hd}{\mbox{H$\delta$}}
\newcommand{\Hg}{\mbox{H$\gamma$}}

\newcommand{\ergscms}{ergs cm$^{-2}$ s$^{-1}$}

\newcommand{\br}{\mbox{$B\!-\!R$}}

\newcommand{\kmsec}{km s$^{-1}$}

\newcommand{\chbeta}{\mbox{$C_{\mathrm{H}\beta}$}}

\newcommand{\msun}{\mbox{$M_\sun$}}

\newfont{\vssn}{cmss10 scaled 1200}

\begin{document}

\received{2008 November 7}

\slugcomment{Submitted for publication in {\it The Astrophysical Journal}}

\title{New Light in Star-Forming Dwarf Galaxies:
The PMAS Integral Field View of the Blue Compact Dwarf Galaxy 
Mrk~409\footnote{Based 
on observations obtained at the German-Spanish
Astronomical Center, Calar Alto, operated by the Max-Planck-Institut f{\"u}r 
Astronomie Heidelberg jointly with the Spanish National Commission for 
Astronomy.}}

\author{Luz M. Cair{\'o}s}
\email{luzma@aip.de}
\affil{Astrophysikalisches Institut Potsdam, An der Sternwarte 16, D-14482 
       Potsdam, Germany} 

\author{Nicola Caon}
\email{nicola.caon@iac.es}
\affil{Instituto de Astrof{\'\i}sica de Canarias, E-38200 La Laguna, Tenerife,
       Spain and Departamento de Astrof{\'\i}sica, Universidad
       de la Laguna, E-38205, La Laguna, Tenerife, Spain}

\author{Polychronis Papaderos}
\email{papaderos@astro.up.pt}
\affil{Centro de Astrof{\'\i}sica da Universidade do Porto, Rua das Estrelas,
4150-762 Porto, Portugal and Instituto de Astrof{\'\i}sica de Andaluc{\'\i}a 
(CSIC), C/ Camino Bajo de Hu{\'e}tor 50, 18008 Granada, Spain}

\author{Carolina Kehrig}
\email{kehrig@aip.de}
\affil{Astrophysikalisches Institut Potsdam, An der Sternwarte 16, D-14482 
       Potsdam, Germany}
       
\author{Peter Weilbacher}
\email{pweilbacher@aip.de}
\affil{Astrophysikalisches Institut Potsdam, An der Sternwarte 16, D-14482 
       Potsdam, Germany}

\author{Martin M. Roth}
\email{mmroth@aip.de}
\affil{Astrophysikalisches Institut Potsdam, An der Sternwarte 16, D-14482 
       Potsdam, Germany}

\author{Cristina Zurita}
\email{czurita@iac.es}
\affil{Instituto de Astrof{\'\i}sica de Canarias, E-38200 La Laguna, Tenerife,
       Spain and Departamento de Astrof{\'\i}sica, Universidad
       de la Laguna, E-38205, La Laguna, Tenerife, Spain}

\accepted{}
\shortauthors{Cair{\'o}s et al.}
\shorttitle{Integral Field Unit Observations of Mrk~409}

\begin{abstract} 

We present an integral field spectroscopic study of the central $2\times2$
kpc$^2$ of the blue compact dwarf galaxy Mrk~409, observed with the 
\emph{Potsdam MultiAperture Spectrophotometer}. This study focuses
on the morphology, two-dimensional chemical abundance pattern, excitation
properties and kinematics of the ionized interstellar medium in the starburst
component. We also investigate the nature of the extended ring of ionized gas
emission surrounding the bright nuclear starburst region of Mrk~409. PMAS
spectra of selected regions along the ring, interpreted with  evolutionary and
population synthesis models, indicate that their ionized emission is mainly
due to a young stellar population with a total mass of $\sim 1.5 \times 10^6$
\msun, which started forming almost coevally $\sim 10$ Myr ago. This stellar
component is likely confined to the collisional interface of a spherically
expanding, starburst-driven super-bubble with denser, swept-up ambient gas,
$\sim 600$ pc away from the central starburst nucleus. The spectroscopic
properties of the latter imply a large extinction ($\chbeta>0.9$), and the
presence of an additional non-thermal ionization source, most likely a
low-luminosity Active Galactic Nucleus. Mrk~409 shows a relatively large
oxygen abundance ($12+\log(\mathrm{O/H})\sim8.4$) and no chemical abundance
gradients out to $R\sim 600$ pc. The ionized gas kinematics
displays an overall regular rotation on a northwest-southwest axis, with a 
maximum velocity of 60 \kmsec; the total mass inside the star-forming 
ring is about $1.4 \times 10^9 M_\sun$. \end{abstract}


\section{Introduction}
\label{Sect:Introduction}

Blue Compact Dwarf (BCD) galaxies are systems undergoing violent bursts of
star formation \citep{Sargent1970}, that appear compact in the optical
(starburst diameter $\leq 1$ kpc), and have low intrinsic luminosities ($M_{B}
\geq -18$ mag) and low gas-phase metallicities ($7.1 \leq
12+\log\mathrm{(O/H)} \la8.3$). These characteristics make them excellent
nearby laboratories for investigating some of the most outstanding questions
in the contemporary extragalactic astronomy: i)~they are suitable sites to
study the process of galaxy formation and evolution, as they are thought to be
the local analogs of the building blocks from which larger systems were formed
at high redshift \citep{Kauffmann1993}; ii)~they allow the determination of
the primordial helium abundance with a minimum of extrapolation to early
conditions \citep{Pagel1992,IzotovThuan2007}; and, iii)~being smaller and less
massive than normal galaxies, BCDs neither can sustain spiral density waves,
nor suffer from disk instabilities, hence they permit the investigation of 
the star-formation process in a relatively simple environment.  

In order to properly characterize the BCD galaxy class, and to get insights
into the above topics, detailed spectrophotometric analysis of individual
objects, aimed to disentangle their different stellar populations and to
elucidate their star-forming histories (SFH), are imperative \citep[see,
e.g.,][]{Cairos2002,Cairos2007}. However, very few such studies have been done 
so far
\citep{Papaderos2002,Papaderos2006,Noeske2000,GildePaz2000a,Fricke2001,
Guseva2001,Cairos2002,Cairos2007}.
These works, all based on traditional observing techniques (broad- and
narrow-band imaging plus long-slit spectroscopy), have shown that combining
spectroscopy and surface photometry is indeed essential for tackling the
problem of BCD evolution. However, they suffer from several drawbacks, the
most severe being the large amount of observing time that they require: 
acquiring images in several broad- and narrow-band filters, in addition to a
series of long-slit exposures sweeping the region of interest, translate into
prohibitively long observing times of two or more nights per galaxy. This
usually means that the data are taken under varying atmospheric conditions
and, potentially, slight differences in the instrumental setup. Clearly,
uncertainties in the combination of such a data set will propagate throughout
in the data analysis and interpretation. Long-slit spectroscopy has also the
additional problem of the uncertainty on the exact location of the slit.

The relatively new observational method of Integral Field Spectroscopy (IFS)
overcomes these problems. Detailed IFS studies have been carried out for
SBS~0335-052\,E with VLT/GIRAFFE \citep{Izotov2006}, for the low-luminosity
BCD II~Zw~70 with CAHA 3.5m/PMAS \citep{Kehrig2008}, for five luminous BCDs
with WHT/INTEGRAL \citep{GarciaLorenzo2008}, for UM~408 with GMOS-IFU 
\citep{Lagos2009} and for Mrk~1418 with PMAS \citep{Cairos2009}.  Although the
scope of these studies is somehow limited by a small field of view (FOV) or by
low spectral resolution, they convincingly demonstrate that IFS is the method
of choice for small, yet complex star-forming (SF) objects like BCDs. Each
single exposure of an integral field spectrograph provides both spatial and
spectral information, making the observations an order of magnitude more
efficient than any  traditional observing technique. Furthermore, IFS provides
simultaneously spectra for all spatial resolution elements, under the same
instrumental and atmospheric conditions, which guarantees the homogeneity of
the dataset. 

Systems with a high degree of asymmetry in their SF component, such as BCDs,
hosting two or more well-separated SF regions, can be studied in a
comprehensive manner only with 2D spectroscopy. Long standing questions in the
field, as the propagation of the star-formation, the generation and dispersal
of heavy elements or the abundance gradients, can be approached efficiently
only with IFS instruments. 

While existing works indicate that the SF activity in BCDs proceeds largely in
bursts \citep[e.g.,][]{Krueger1994,MasHesse1999}, we are still lacking an
understanding of the two-dimensional star-formation pattern. Established is
solely that, in the majority (90\%) of BCDs, the SF regions are not randomly
scattered within the low surface brightness (LSB) host, but rather confined to
its central part, out to $\sim 2$ exponential scale lengths
\citep{Papaderos1996,Cairos2001}. This fact suggests a physical link between
the mode of the SF activity in BCDs and the shape of the gravitational
potential formed by the old, higher mass-to-light (M/L) ratio stellar LSB
host.

Some of the most metal-poor BCDs, thought to be still in the process of their
formation, such as SBS~0335-052\,E \citep{Thuanetal1997,Papaderos1998}  or
SBS~1415+437 (\citealt{Thuan1999}; \citealt{Guseva2003}; see
\citealt{Papaderos2008} for a review), present clear signatures of
uni-directional SF propagation.  This process can lead to a \emph{cometary}
morphology, with the ``comets head'' (the dominant SF region) located at the
tip of an elongated LSB host delineating the trail of past propagating SF
activities \citep{Papaderos2008}. An irregular SF propagation pattern in form
of a random-walk was reported  by a detailed analysis of individual stellar
clusters in the luminous BCD ESO~338-IG04 \citep{Ostlin2003}, while the
potential role of SF propagation in Mrk~370 and Mrk~35 has been discussed in 
\cite{Cairos2002} and \cite{Cairos2007}, respectively. 

In the two BCDs Mrk~86 and Mrk~409, the ring-morphology of the SF regions 
suggests that the SF activity could have been triggered by the shock wave
produced by multiple SNs in a central starburst. In the case of Mrk~86,
extensive studies \citep{GildePaz2000b,GildePaz2002} support this hypothesis.
However, because of the small spatial and spectral coverage of the available
spectroscopic observations \cite[][hereafterGdP03]{GildePaz2003a}, conclusive
results for Mrk~409 can not be drawn.

In this paper we show the great potential of IFS when applied to BCDs by
presenting a pilot study focused on the galaxy Mrk~409 (= NGC~3011).  Mrk~409
is a luminous ($M_{B} = -17.73$) BCD, belonging to the nE morphological class 
\citep{LooseThuan1986}. Its intriguing gas ionized morphology --- a compact
nuclear starburst region surrounded by two SF rings at projected
galactocentric radii of 5\arcsec\ and 18\arcsec, corresponding to 0.64 and 2.3
kpc --- makes it an excellent laboratory for investigating the star-formation
pattern and abundance gradients in BCDs. We observed the galaxy with the PMAS
spectrograph mounted on the 3.5m Calar Alto telescope. The PMAS field of view
($2\times2$ kpc$^{2}$ at the Mrk~409 distance) allows us to map the nuclear
region and the inner SF ring. The basic parameters of the galaxy are
summarized in  Table~\ref{Table:data}.

The paper is organized as follows: in Sect. \ref{Sect:Data} we describe the
observations, the data analysis and the method employed to derive the two
dimensional maps. In Sect. \ref{Sect:Results} we present continuum and
emission line intensity, line ratio and velocity maps as well as results from
the analysis of the integrated spectra of the SF regions identified in the
galaxy. The results are discussed and summarized in 
Sect.~\ref{Sect:Discussion} and Sect.~\ref{Sect:Conclusions}, respectively.


\begin{deluxetable}{lccccccc}
\footnotesize
\tablewidth{0pt}
\tablecaption{Basic data for Mrk~409}
\tablehead{
  \colhead{Parameter}  & \colhead {Data} & \colhead {Note} }
\startdata
 Other names                        &  NGC~3011, UGC~5259    	   &      \\
 RA~(2000)                          &  \phs09 49 41	    	   &      \\
 DEC~(2000)                         &  +32 13 16 	    	   &      \\
 $v_\mathrm{helio}$~(\kmsec)        &  1527		    	   &      \\
 $D$ (Mpc)                          &  26.3		    	   &	  \\
 $A_{B}$~(mag)                      &  0.071	                   &  (1) \\
 $M_{B}$~(mag)                      &  $-17.73$  	           &      \\
 $m_{B}$~(mag)                      &  $14.37\pm0.03$	           &  (2) \\
 $m_{R}$~(mag)                      &  $13.33\pm0.05$	           &  (2) \\
 $F(\Ha)$~(ergs s$^{-1}$ cm$^{-2}$) &  $2.93\pm0.17\times10^{-13}$ &  (2) \\
 $m_{J}$~(mag)               	    &  $11.88\pm0.02$		   &  (3) \\
 $m_{H}$~(mag)               	    &  $11.34\pm0.03$		   &  (3) \\
 $m_{K_{s}}$~(mag)           	    &  $11.17\pm0.04$		   &  (3) \\
 $M_{HI}$~($M_\sun$)         	    &  $0.31\times10^{9}$	   &  (4) \\
 $M_\mathrm{T}$~($M_\sun$)   	    &  $0.33\times 10^{10}$	   &  (4) \\
\enddata
\label{Table:data}
\tablecomments{Coordinates, heliocentric velocity and distance, all taken 
from NED (http://nedwww.ipac.caltech.edu/).
Distance calculated using a Hubble constant of 73 \kmsec\ Mpc$^{-1}$, 
and taking into account the influence of the Virgo Cluster, the Great 
Attractor and the Shapley supercluster.
Absolute magnitude in the $B$ band, computed from the tabulated 
integrated $B$ magnitude and distance.
(1) Absorption coefficient in the $B$ band, from \cite{Schlegel1998}.
(2) \cite{GildePaz2003b}.
(3) From 2MASS.
(4) Neutral hydrogen mass $M_{HI}$ and total mass $M_\mathrm{T}$ 
from \cite{ThuanMartin1981}.}
\end{deluxetable}

\section{The Data}
\label{Sect:Data}

\subsection{Observations and Data Reduction} 
\label{SubSect:Observations}


Mrk~409 was observed in 2007 March, with the \emph{Potsdam MultiAperture
Spectrophotometer} (PMAS) attached to the 3.5m telescope of the Observatorio
Astron{\'o}mico Hispano Alem{\'a}n Calar Alto (CAHA). PMAS in an integral
field spectrograph, with a lens array of $16''\times 16''$ ($2 \times
2$~kpc$^{2}$ at the Mrk~409 distance) square elements. This array is connected
to a bundle of fiber optics, whose 256 fibers are re-arranged to form a
pseudo-slit in the focal plane of the fiber spectrograph. In the configuration
used each fiber corresponds to a spatial sampling on the sky of
$1\arcsec\times 1\arcsec$ (see \citealp{Roth2005,Kelz2006} for more details
about the instrument). A grating with 300 grooves per mm was used during the
observations, in combination with a $2048 \times 4096$ pixel SITe ST002A CCD
detector; this provides a spectral range of 3590--6996 \AA\, with a linear
dispersion of 3.2 \AA\ per pix and a resolution (FWHM) of about 7 \AA.

We observed a total of 6000 sec on the galaxy, with the integration time split
into five exposures of 1200 sec each; additional sky frames were taken  moving
the telescope several arcmin away from the target position. Calibration frames
were observed before and after the galaxy exposures. These calibrations
consist of spectra of emission line lamps (HgNe lamp), which are required to
perform the wavelength calibration, and spectra of a continuum lamp, necessary
to locate the fiber spectra on the CCD and to perform the flat-fielding
correction. The spectrophotometric standard stars BD+75325 and BD+332642 were
observed for flux calibration. As usual, bias and sky-flat exposures were
taken at the beginning and at the end of the night. The seeing was about 1.8
arcsec.

The data were processed using standard IRAF\footnote{IRAF is distributed
by the National Optical Astronomy Observatories, which are operated by the
Association of Universities for Research in Astronomy, Inc., under cooperative
agreement with the National Science Foundation.} tasks. The reduction
procedure included bias subtraction and image trimming, tracing and
extraction, wavelength and distortion calibration, flat-fielding, combination
of the individual frames, sky subtraction and flux calibration. For a detailed
description of the individual steps performed in fiber data processing see
\cite{Cairos2009}. The typical errors on the wavelength calibration are  
0.02 \AA. 
The flux calibration was carried out using our observations of 
spectrophotometric standard stars. By comparing the sensitivity curves 
obtained for all the spectrophotometric standards observed throughout the 
three nights of the observing run, we estimate that the relative uncertainty 
on the calibration factor is generally equal or less than 2\%, except blueward 
of 4000 \AA, where the curves show a marked change of slope and the 
uncertainty increases up to about 8\%.

\subsection{Line Fitting and Map Generation}
\label{SubSect:LineFitting}

In this study we generate 2D intensity maps for the most important
emission lines as well as for different continuum regions.  The relevant
parameters of the emission lines (center, flux and equivalent width) were
determined by fitting single Gaussians; the fit was carried out by using the
$\chi^{2}$ minimization algorithm implemented by C.~B. Markwardt in the
\emph{mpfitexpr} IDL library \citep{Markwardt2009}.
The \Hb\ line, in which the absorption component was also visible, was modeled 
as the sum of two Gaussians. The continuum (typically 30--50 \AA\
on both sides) was fitted by a straight line. Lines in a doublet were fitted
imposing that they have the same redshift and width.

Criteria such as flux, error on flux, velocity and width were used to do a
first, automatic assessment of whether to accept or reject a fit. For
instance, lines with too small (less than the instrumental width) or too large
($> 5$ \AA) widths were flagged out, similar to lines with an error on the
flux of more than about 10\% (the exact limits depending on the specific
line). In order to improve the reliability and stability of fits invoking both
an absorption and emission component, an additional criterion was applied to
the equivalent width of the absorption component (allowed to vary between 0
and 5 \AA). 

Each individual fit was visually checked, and when necessary, the
aforementioned automated quality criteria were overridden (see
\citealt{Cairos2009}).

The fit procedure gives, for each spectral line, a table with the spaxel 
number and the values of the measured parameter; this table is then used to
build a 2D spatial map. 
Continuum maps were built in the same way by summing the signal within
wavelength intervals free from emission lines or strong sky-lines residuals.

\section{Results}
\label{Sect:Results}

\subsection{Continuum and Emission Line Intensity Maps}
\label{SubSect:LineMaps}

\begin{figure}   
\begin{center}
\includegraphics[width=\textwidth]{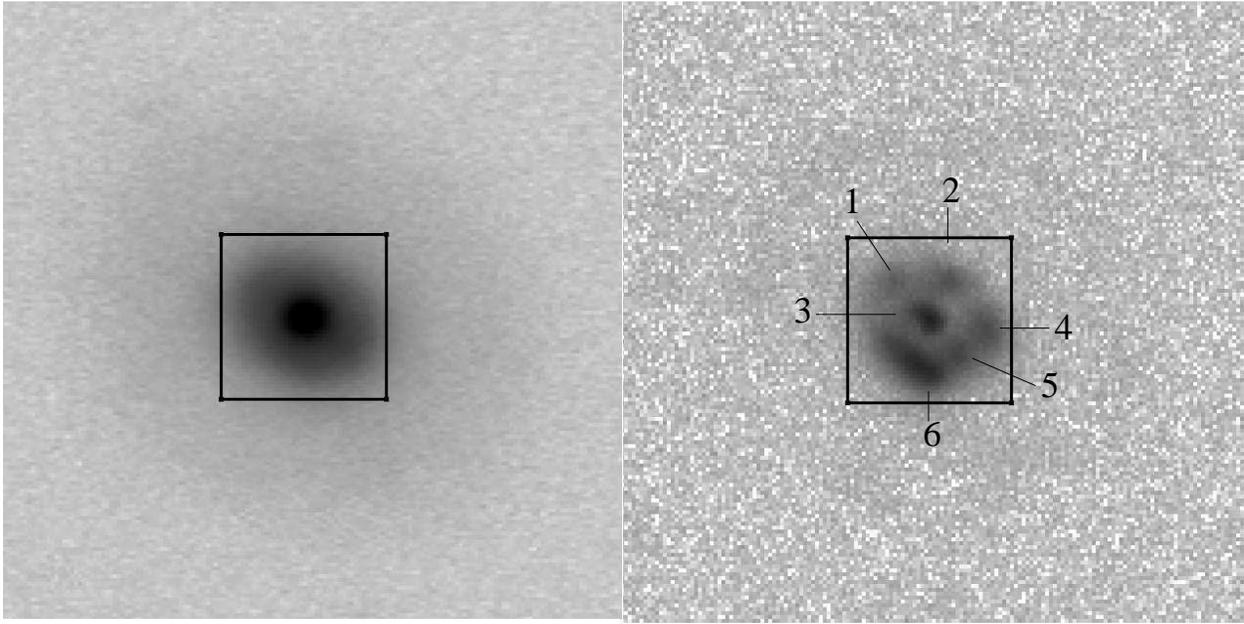}
\caption{\emph{left panel}: B-band image of Mrk~409; the central box indicates
the field of view covered by our PMAS observations; \emph{right panel}: 
Continuum-subtracted \Ha\ image with the individual SF knots labeled. 
The field of view is one arcmin in both frames. 
North is up, east to the left. Both images, retrieved from the NED, are 
published in \cite{GildePaz2003b}}
\label{Fig:images}
\end{center}
\end{figure}

The main advantage of IFS is that it maps an object simultaneously in
different continuum bands and emission lines, so that we can obtain at the
same time the spatial distribution of the stellar component and of the ionized
gas emission. We constructed continuum and flux maps of the brightest
emission lines in Mrk~409. To allow a quick comparison with our results, the 
$B$-band and \Ha\ images of the galaxy \citep{GildePaz2003b} are displayed in
Figure~\ref{Fig:images}.

\begin{figure*}
\mbox{
\centerline{
\hspace*{0.0cm}\subfigure{\includegraphics[width=5.5cm]{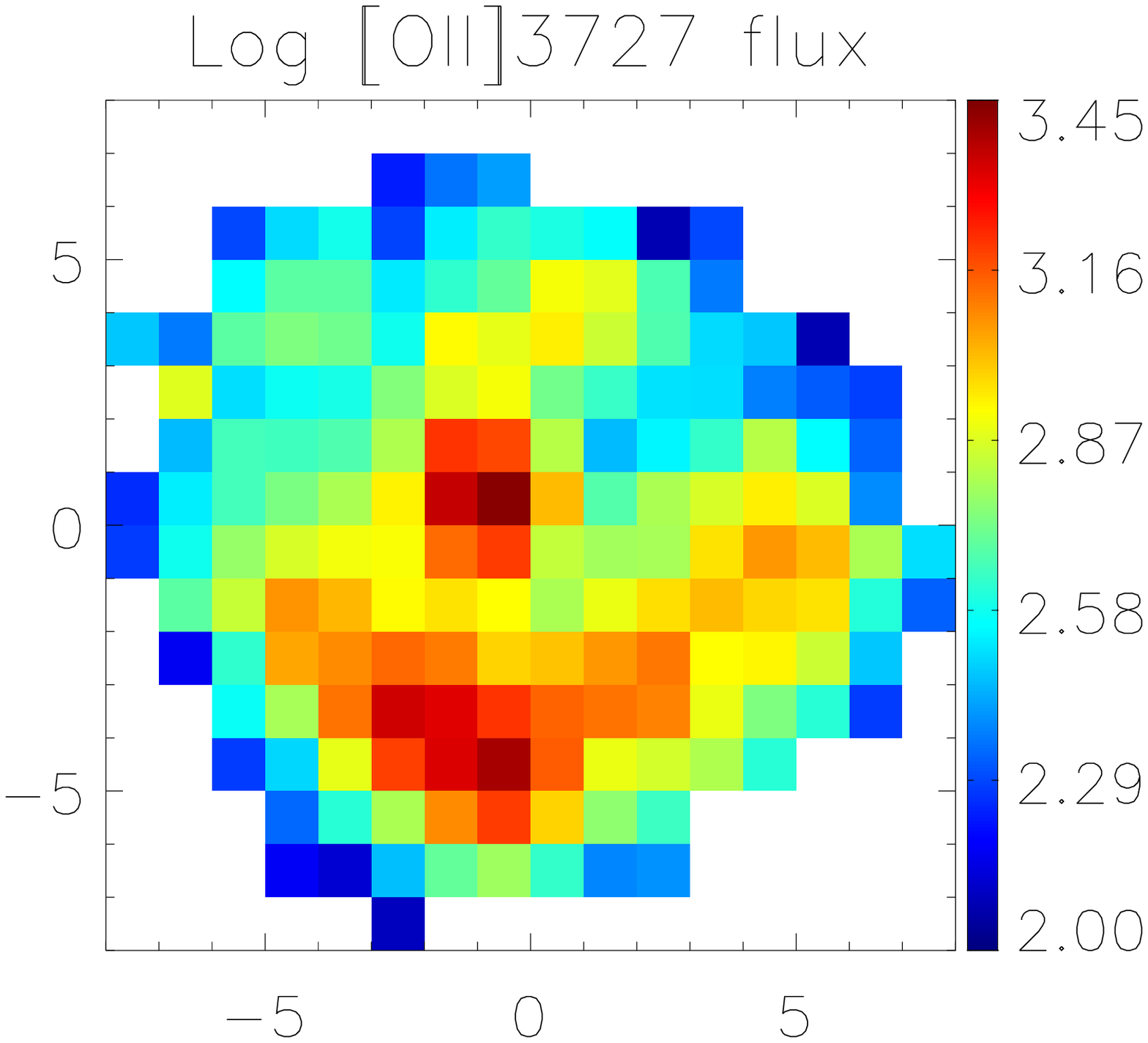}}
\hspace*{0.0cm}\subfigure{\includegraphics[width=5.5cm]{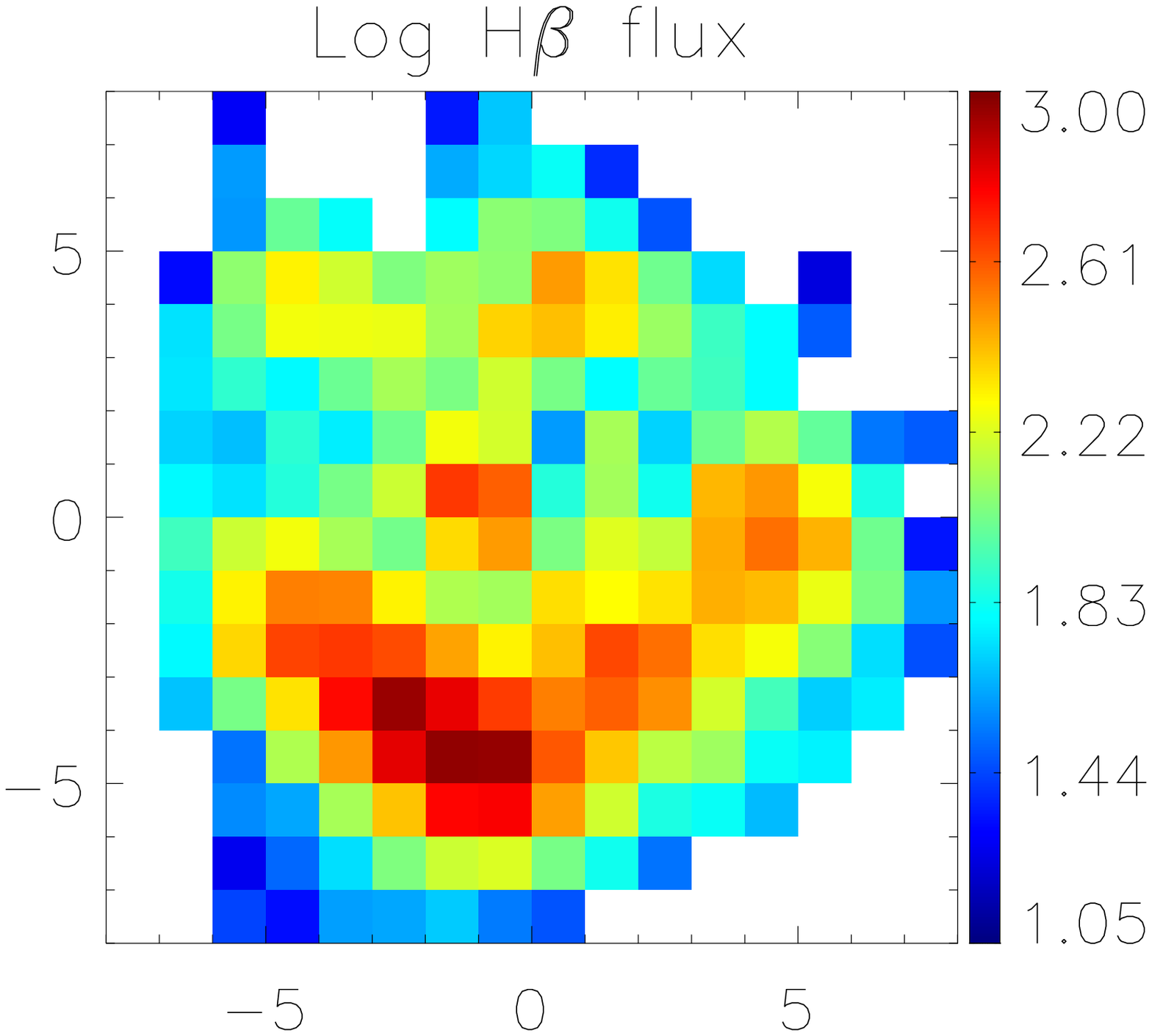}}
\hspace*{0.0cm}\subfigure{\includegraphics[width=5.5cm]{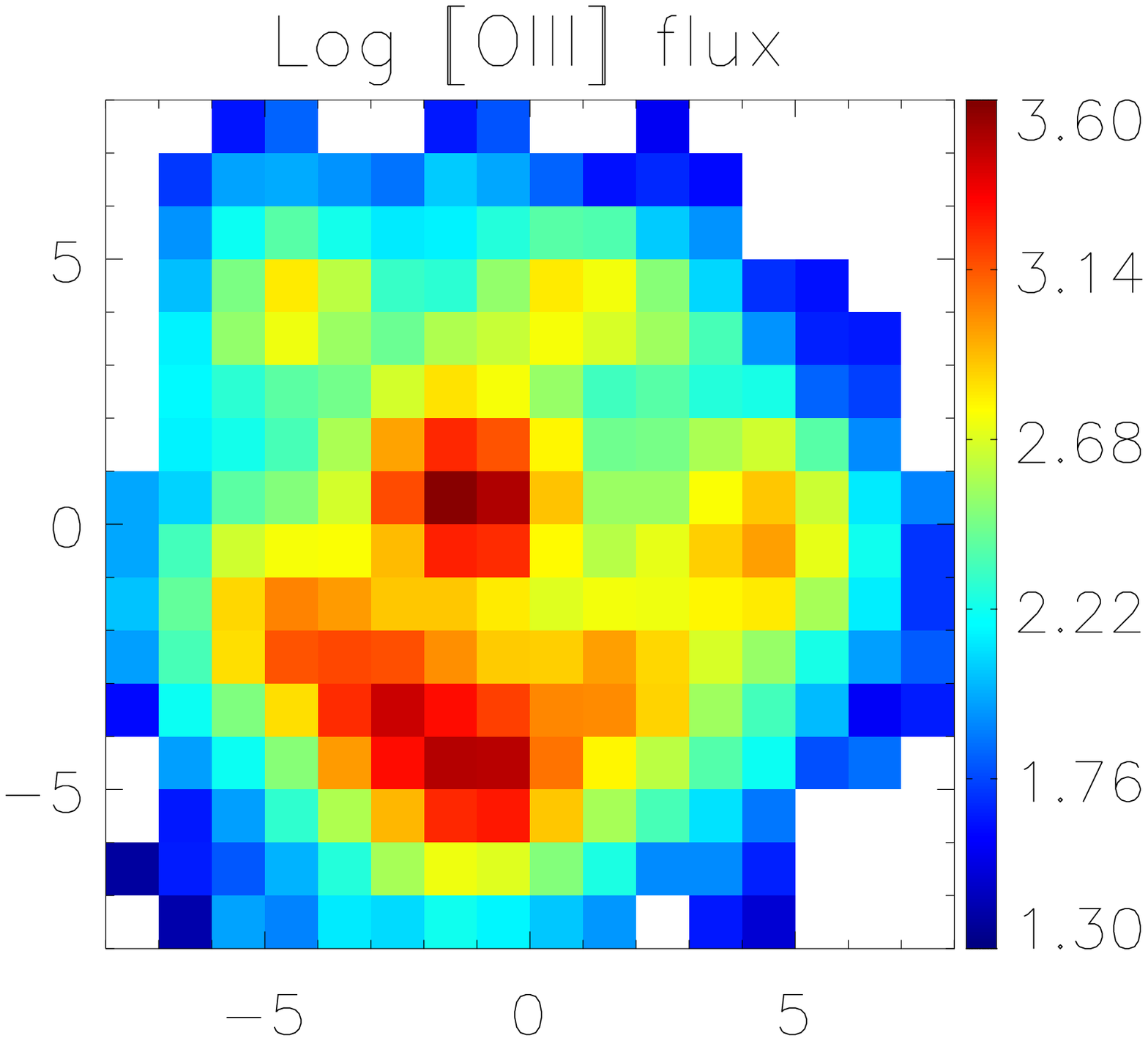}}
}}   
\mbox{
\centerline{
\hspace*{0.0cm}\subfigure{\includegraphics[width=5.5cm]{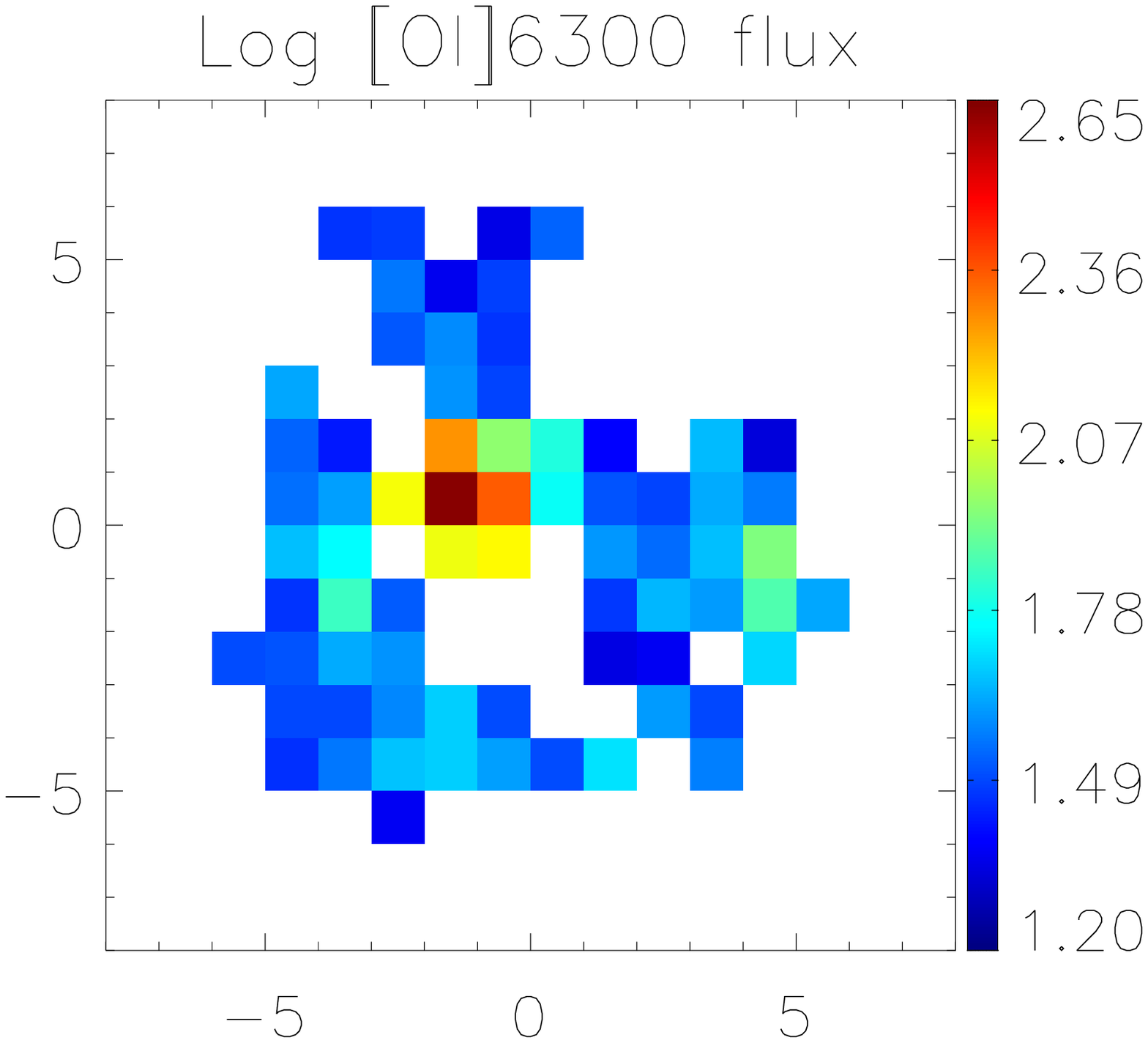}}
\hspace*{0.0cm}\subfigure{\includegraphics[width=5.5cm]{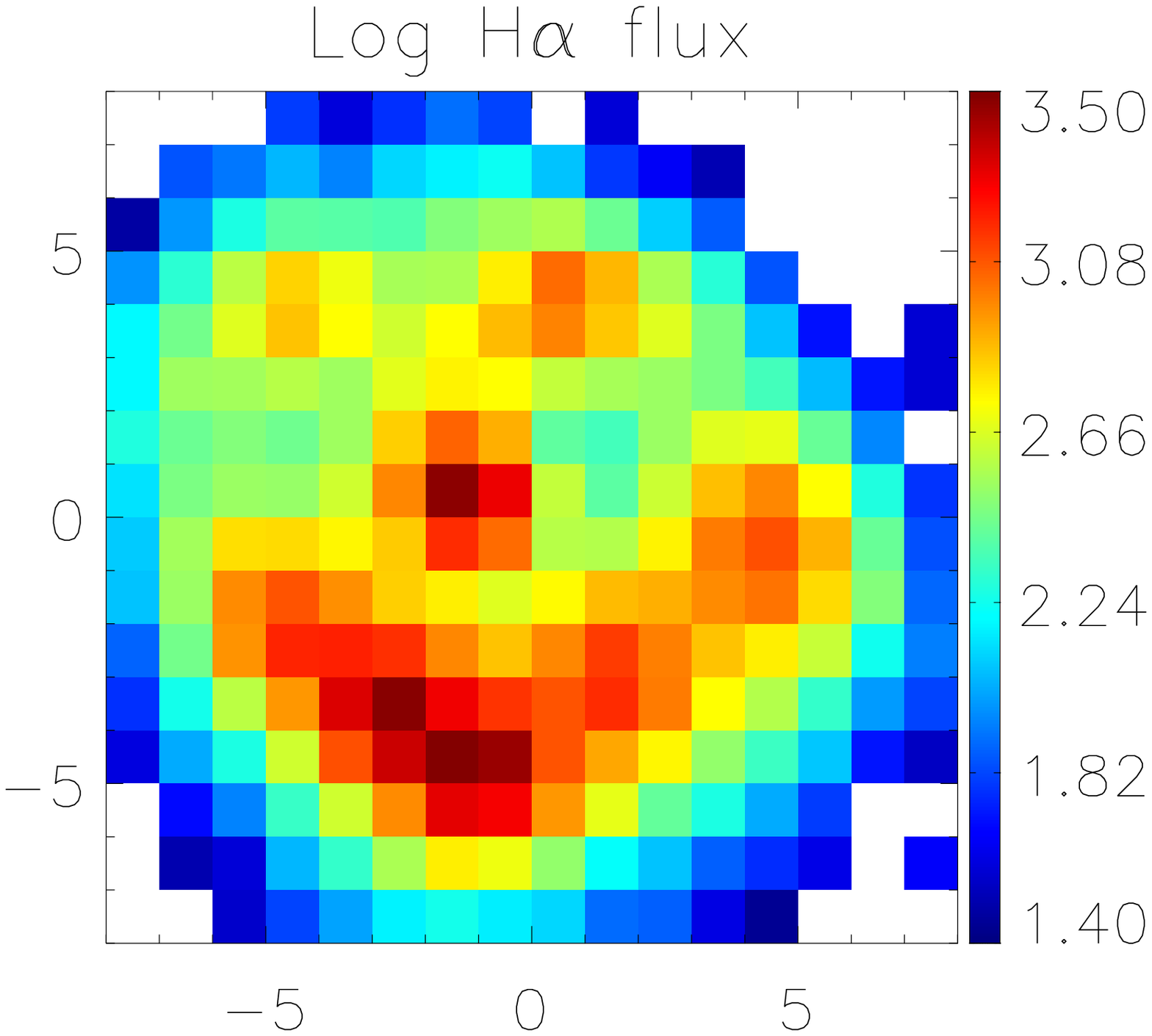}}
\hspace*{0.0cm}\subfigure{\includegraphics[width=5.5cm]{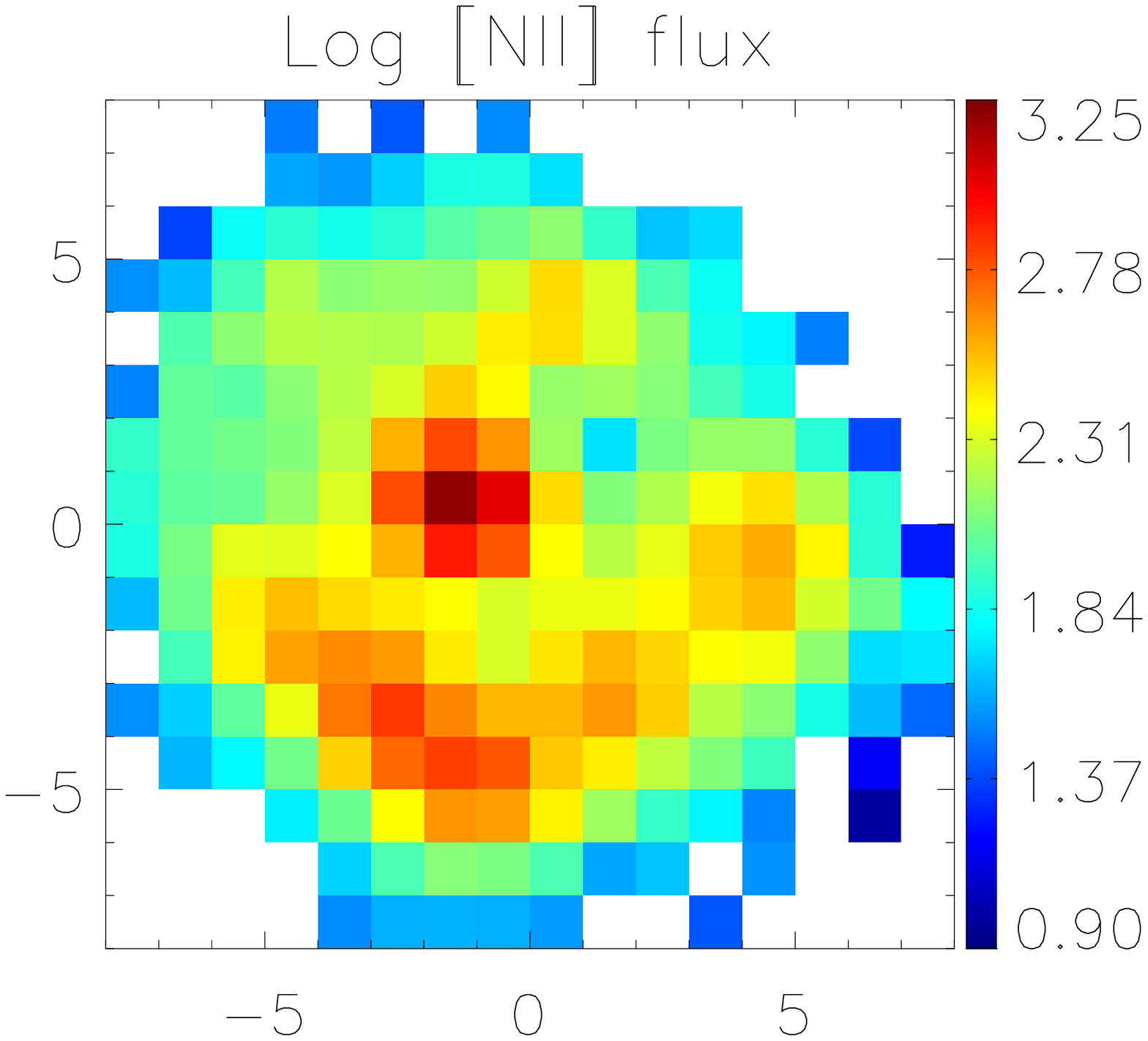}}
}}   
\mbox{
\hspace*{0.0cm}\subfigure{\includegraphics[width=5.5cm]{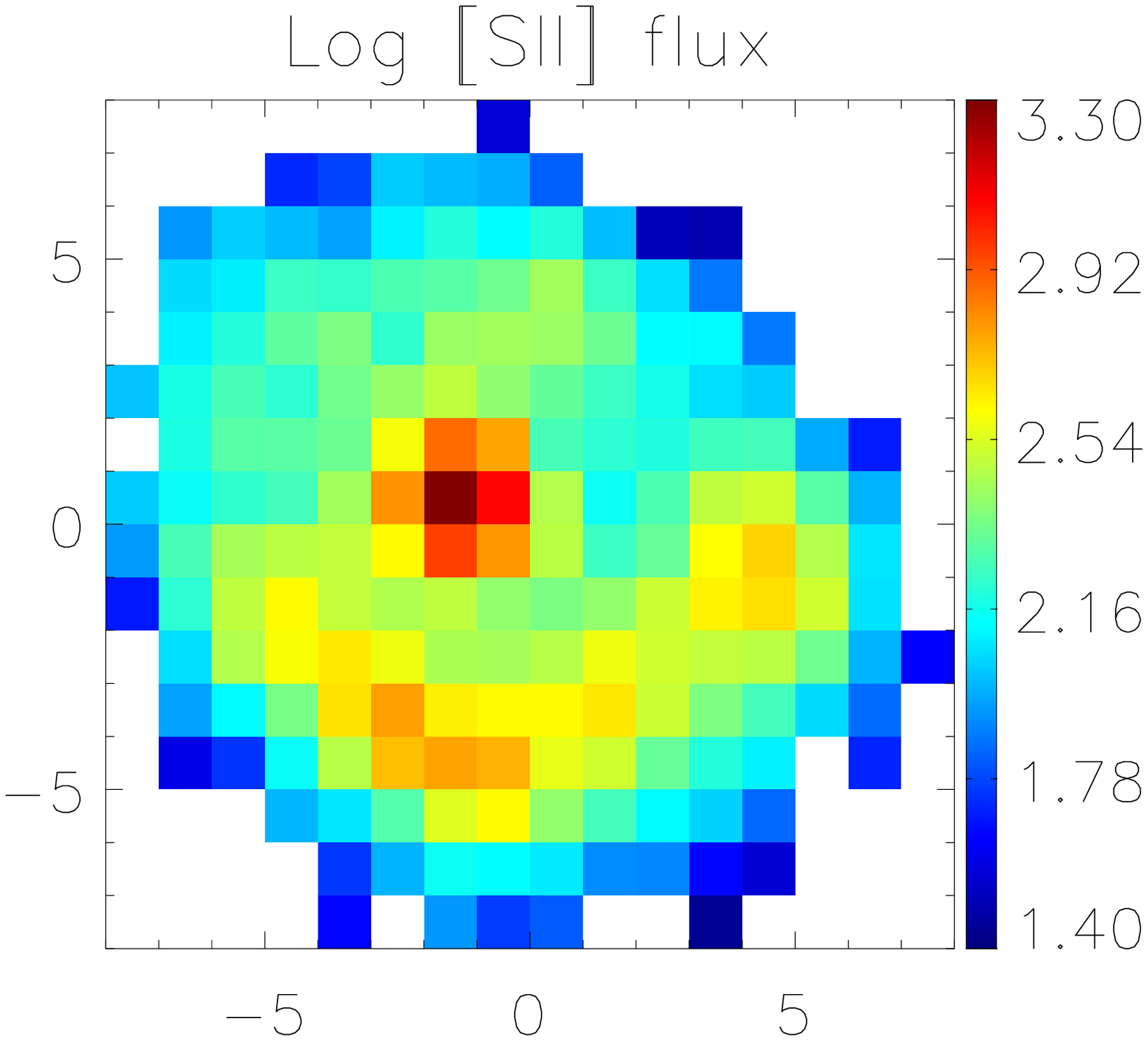}}
\hspace*{0.0cm}\subfigure{\includegraphics[width=5.5cm]{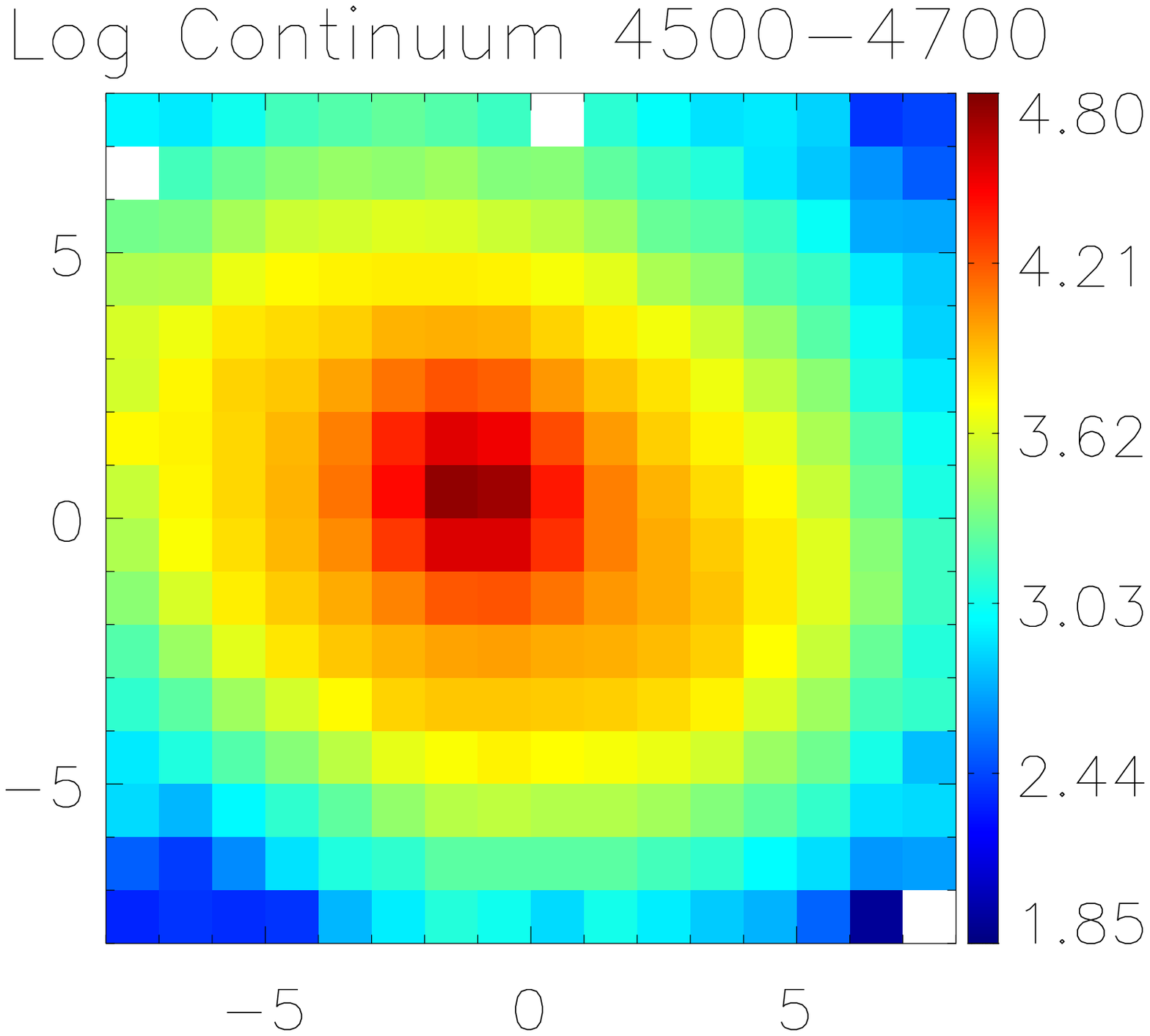}}
}   
\caption{Emission line intensity maps derived from Gaussian fits 
to each line and spaxel (see text for details):
[\ion{O}{2}]~$\lambda3727$; 
\Hb; 
[\ion{O}{3}]~$\lambda5007$;
[\ion{O}{1}]~$\lambda6300$; 
\Ha; 
[\ion{O}{3}]~$\lambda5007$;
[\ion{S}{2}]~$\lambda6717,\;6731$. 
The last panel shows the continuum emission between 4500 and 4700 \AA,  
a spectral region free from line emission (``pure continuum'').
North is up, east to the left. Axis units are arcseconds. 
All images are shown in logarithmic scale to bring up the fainter 
regions. Flux units are $10^{-18}$ ergs cm$^{-2}$ s$^{-1}$ \AA$^{-1}$.}
\label{Fig:linemaps}
\end{figure*}

Figure~\ref{Fig:linemaps}  displays  the   [\ion{O}{2}]~$\lambda3727$,  \Hb,
[\ion{O}{3}]~$\lambda5007$, [\ion{O}{1}]~$\lambda6300$, \Ha,
[\ion{N}{2}]~$\lambda6584$  and  [\ion{S}{2}]~$\lambda\lambda6717,\;6731$
emission-line   maps.  These maps  trace  the  regions of  recent  star
formation. In all maps except one Mrk~409 shows a similar, irregular pattern,
with a number of SF knots distributed in a ring of diameter  $\simeq 5$ arcsec
(600 pc) around the nuclear starburst (region \#3); the ring is resolved into
at least five smaller SF knots. Only in [\ion{O}{1}]~$\lambda6300$ the galaxy
has a strikingly different morphology, revealing solely a single prominent
central peak.  The presence of strong [\ion{O}{1}]~$\lambda6300$ is an 
indicator that power-law photoionization or shock-heating are important in the
nuclear region of Mrk~409.

The stellar continuum emission in the spectral range 4500--4700 \AA\ is also
shown in Figure~\ref{Fig:linemaps}. This map displays a regular morphology
with elliptical isophotes with nearly constant position angle, and a well
defined central peak, whose location is the same in all continuum bands and
also coincides with the nuclear SF region \#3.

\subsection{Line Ratio Maps} 
\label{SubSect:LineRatios}

\begin{figure*}
\mbox{
\centerline{
\hspace*{0.0cm}\subfigure{\includegraphics[width=5.5cm]{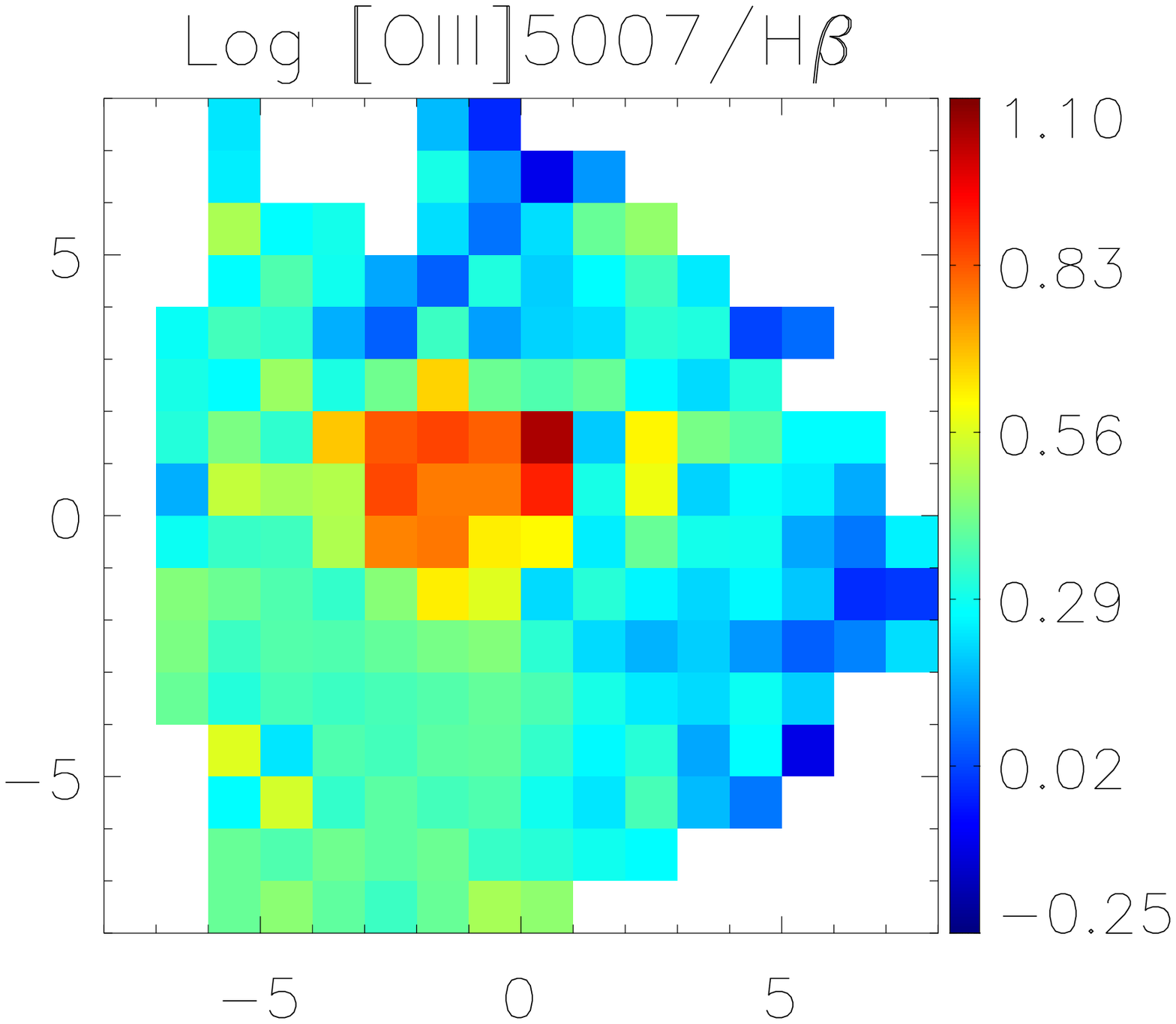}}
\hspace*{0.0cm}\subfigure{\includegraphics[width=5.5cm]{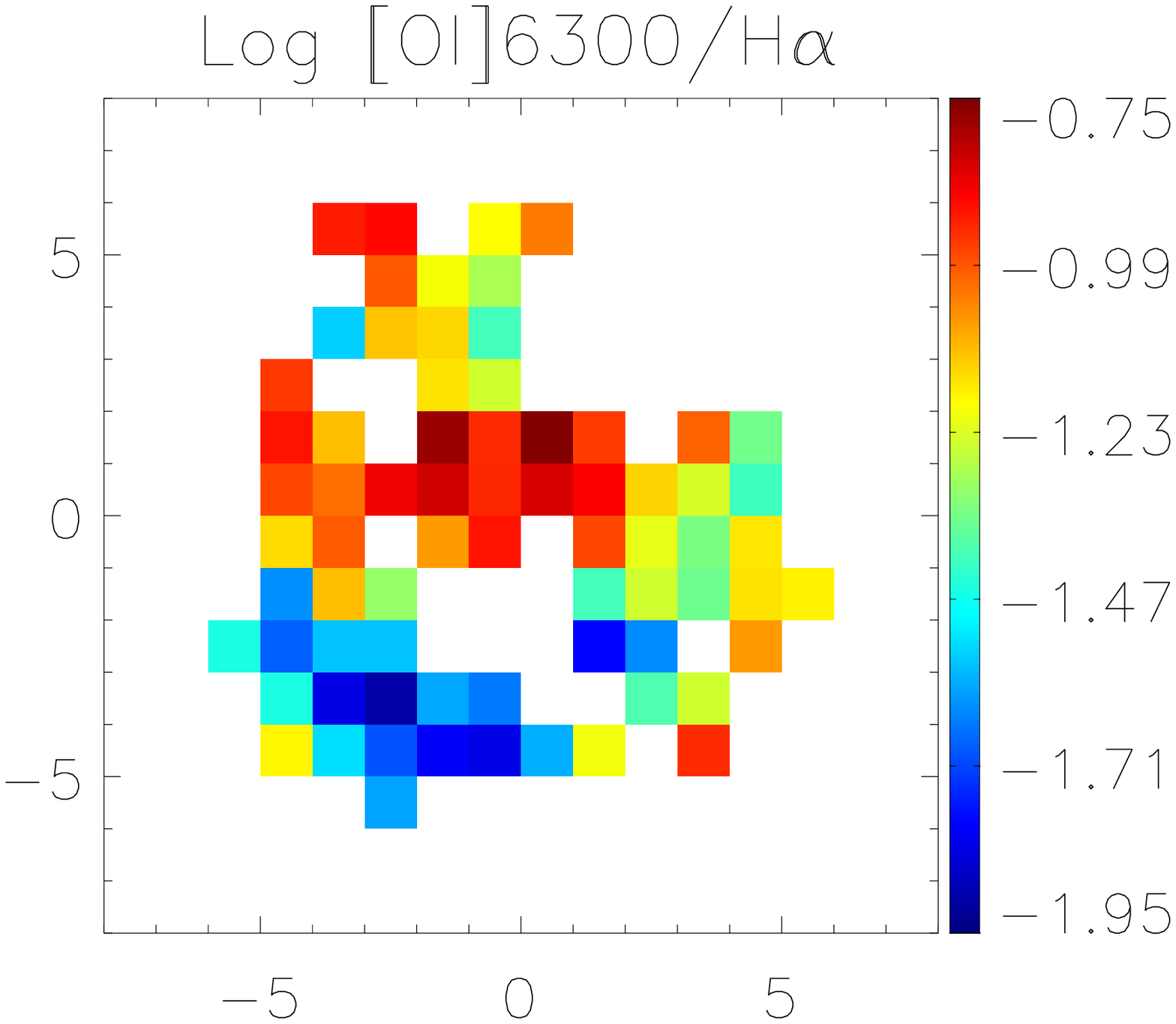}}
\hspace*{0.0cm}\subfigure{\includegraphics[width=5.5cm]{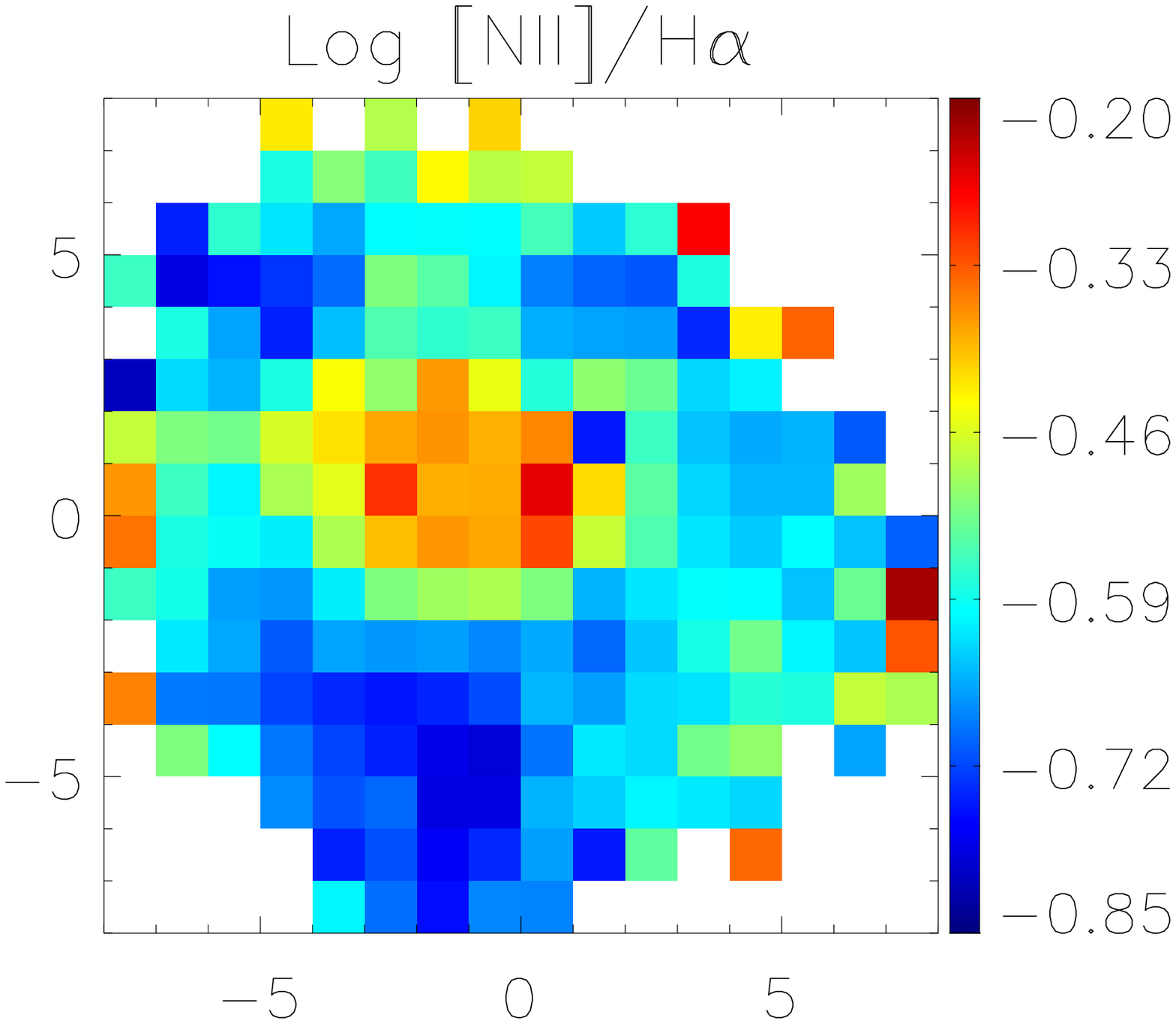}}
}}   
\mbox{
\centerline{
\hspace*{0.0cm}\subfigure{\includegraphics[width=5.5cm]{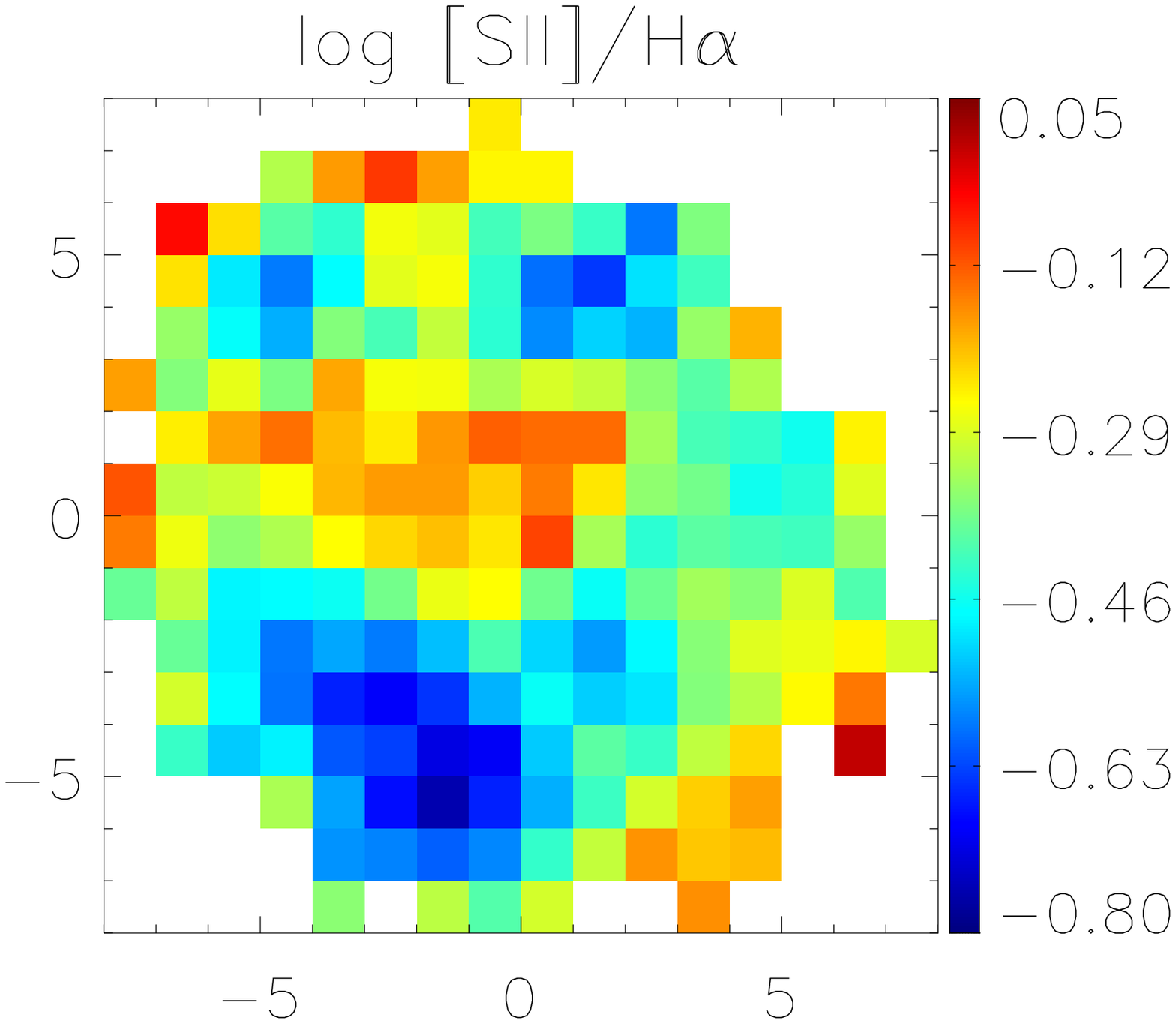}}
\hspace*{0.0cm}\subfigure{\includegraphics[width=5.5cm]{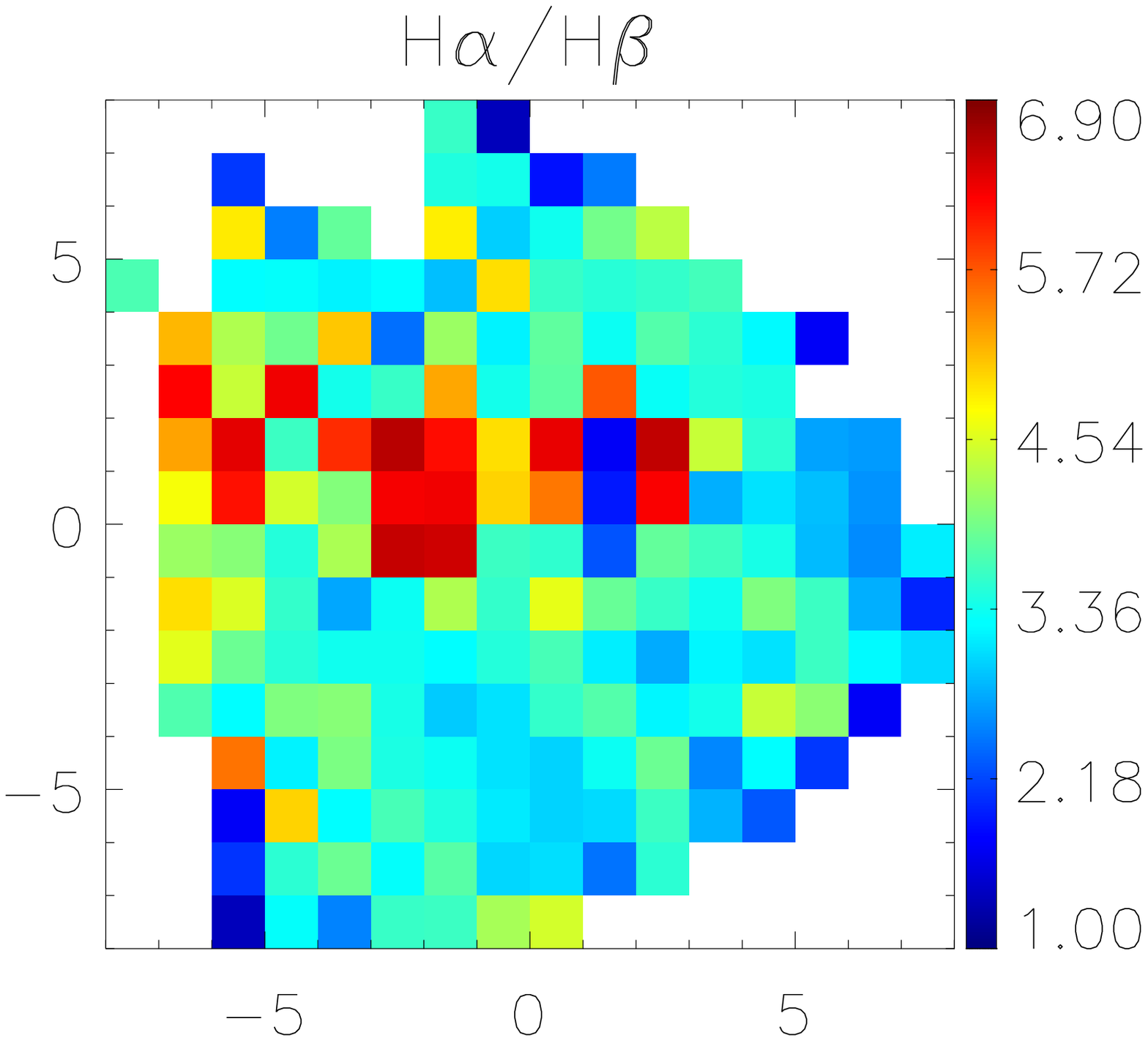}}
\hspace*{0.0cm}\subfigure{\includegraphics[width=5.5cm]{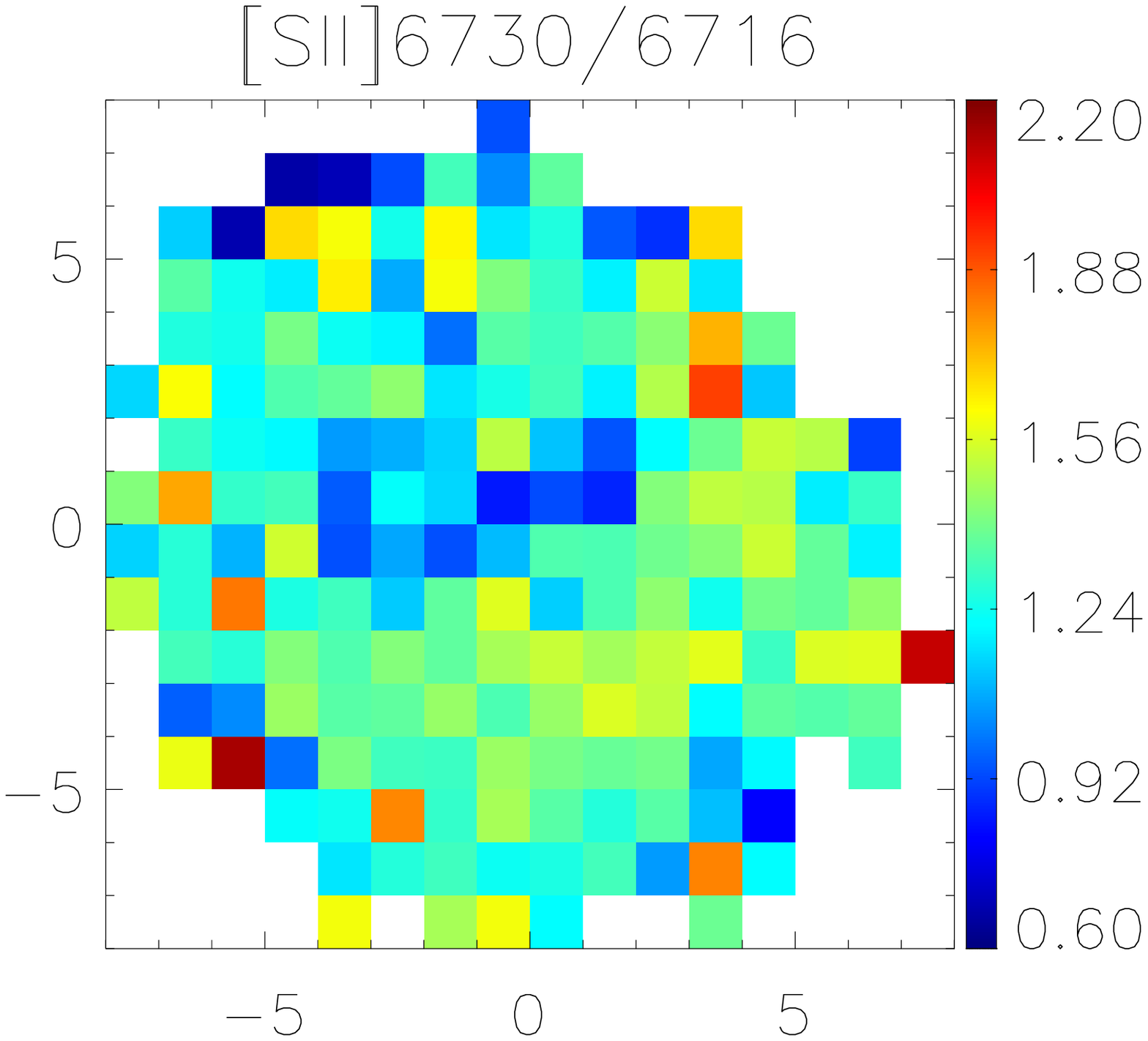}}
}}   
\caption{Ionization ratios (first four panels), interstellar extinction 
 and electron density  maps. Axis units are 
arcseconds; north is up, east to the left. Ionization ratio maps are in 
logarithmic scale.}
\label{Fig:lineratios}
\end{figure*}

In order to investigate the main ionization mechanisms operating in Mrk~409, we
built the [\ion{O}{3}]~$\lambda5007$/\Hb, [\ion{O}{1}]~$\lambda6300$/\Ha,
[\ion{N}{2}]~$\lambda6584$/\Ha\ and
[\ion{S}{2}]~$\lambda\lambda6717,\;6731$/\Ha\ line ratio maps, shown  in
Figure~\ref{Fig:lineratios}.  In these maps high excitation corresponds to high
values of [\ion{O}{3}]~$\lambda5007$/\Hb, and to low values of
[\ion{N}{2}]~$\lambda6584$/\Ha\ and
[\ion{S}{2}]~$\lambda\lambda6717,\;6731$/\Ha.  All the excitation maps display a
similar pattern. In the SF knots in the ring, the values of the different ratios
are those characteristic of  \ion{H}{2} regions. In the central starburst,
however, opposite to what is expected in regions  photoionized by stars, high
values for all ratios are found: [\ion{O}{3}]~$\lambda5007$/\Hb $\geq 6$; 
[\ion{O}{1}]~$\lambda6300$/\Ha\ $> 0.2$;  [\ion{N}{2}]~$\lambda6584$/\Ha\ $>
0.5$;  [\ion{S}{2}]~$\lambda\lambda6717,\;6731$/\Ha\ $> 1$), most probably
indicating the presence of an additional ionization mechanism  --- an
\textsl{Active Galactic Nucleus} (AGN) or shocks.

The extinction in the galaxy (Figure~\ref{Fig:lineratios}) has been derived from
the \Ha/\Hb\  ratio. Mrk~409 displays an inhomogeneous extinction pattern, which
peaks in  the nuclear starburst. In this region the \Ha/\Hb\ line ratio rises to
values as high as 6, suggesting strong extinction by dust with a reddening 
$E(B-V)$ of up to  $\simeq 0.70$ mag. By contrast, in the circumnuclear SF knots
the \Ha/\Hb\ ratio is only  slightly larger than the nominal value of 2.86
\citep{Osterbrock2006}, and  implies $E(B-V)\la0.2$.  Note that, applying the
extinction derived in the central region  to the whole SF component, as is
usually done in long-slit or  single-aperture spectroscopic studies (with, e.g.,
the SDSS), would   lead to a large overestimate of the \Ha\ luminosity and
consequently of the  Star Formation Rate (SFR). In this respect, a spatially
resolved 2D determination of the extinction is  essential for a proper
characterization of the stellar populations and the  star formation history of
Mrk~409, and more in general of BCD galaxies.

In computing the extinction map, we corrected the \Hb\ line for underlying
stellar absorption by fitting a Gaussian profile to its absorption wings. For
the \Ha\ line, the absence of absorption wings makes a reliable decomposition
impossible. In this case, in principle several approaches could be adopted : a)
to take for \Ha\ the same value in absorption that for \Hb,  that is, to set 
$\mathrm{EW}(\Ha_\mathrm{abs})=\mathrm{EW}(\Hb_\mathrm{abs})$, traditionally 
the most common strategy;  b) to apply spectral synthesis models to disentangle
the young and old  population spectra and, from the model output, estimate the
EW(\Ha) in  absorption; c) to assume that the absorption in \Ha\ is negligible.

Taking into account that both a) and b) rest on "model-dependent" values,
probably overestimations of the underlying absorption in \Ha, which in turn
overestimates \chbeta, we prefer to adopt the most conservative c) approach,
keeping in mind that  under this assumption the computed extinction value is a
lower limit to the actual value.

\subsection{Integrated spectroscopy of selected galaxy regions}
\label{SubSect:IntegratedSpectroscopy}

\begin{figure}   
\begin{center}
\includegraphics[width=0.7\textwidth,angle=-90]{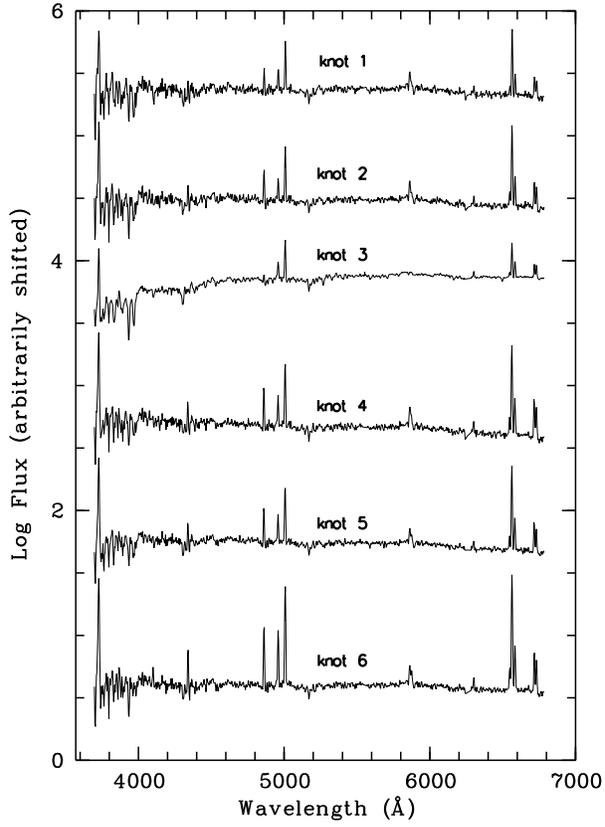}
\caption{Flux-calibrated spectra of the six resolved star-forming regions 
in Mrk~409. The spectra are shown in logarithmic scale and are offset for  
the sake of clarity.}
\label{Fig:spectra}
\end{center}
\end{figure}

As the next step in our analysis, we used  the reconstructed flux and
continuum maps to delimit the regions of interest in the galaxy.  A total of
six regions were identified, the five SF located in the ring and the nuclear
SF region \#3 (see Fig.~\ref{Fig:images}).

We identified the spaxels corresponding to each of these regions, and summed
their spectra together to create the integrated spectrum. In this way we
produce higher signal-to-noise spectra (in comparison to the spectra of the
individual spaxels), better suited for the derivation of physical parameters
and chemical abundances. 

In Figure~\ref{Fig:spectra} we compare the flux-calibrated spectra of these
regions, prior to correction for interstellar extinction. The spectra are
arbitrarily shifted on the vertical axis, for the sake of clarity. It can be
seen that the spectral energy distribution (SED) of the nucleus  \#3 differs
strikingly from that in regions along the circumnuclear SF ring (\#1, \#2,
\#4--6). In the nuclear starburst the stellar continuum is an order of
magnitude higher than in the ring; it displays prominent absorption features,
including the higher-order Balmer lines, \Hd\ and \Hg, CaII~H and K lines,
NaI~$\lambda5922$ or the G-band at $\lambda$4304. Its strong
[\ion{O}{1}]~$\lambda6300$ emission is also remarkable, suggesting an
additional contribution to the ionization, other than photoionization from OB
stars. This comparatively red stellar continuum and the strong absorption
features both indicate that the emission is dominated by older stars. The SF
regions in the ring display, on contrast, an almost flat continuum with
stronger emission lines, such as [\ion{O}{2}]~$\lambda3727$, \Hb,
[\ion{O}{3}]~$\lambda5007$, \Ha, [\ion{N}{2}]~$\lambda\lambda6548,\;6584$ and
[\ion{S}{2}]~$\lambda\lambda6717,\;6731$, in addition to weak
[\ion{O}{1}]~$\lambda6300$ emission. These spectra also reveal absorption
wings in \Hd\ and \Hg\, and such absorption features as CaII~H~K
$\lambda\lambda3933,\;3969$.

We measured the fluxes and equivalent widths of the most prominent emission
lines in each spectrum. The extinction in each region was determined from the
\Ha/\Hb\ ratio, as the measured \Hd\ and \Hg\ fluxes are doubtful due to their
intrinsic weakness and the large uncertainties in the correction for the
underlying stellar absorption. For the theoretical \Ha/\Hb\ value, we adopted
the case B Balmer recombination decrement $\Ha/\Hb=2.86$ for  $T=10000$ K and
$N_\mathrm{e}=10^{4}$ cm$^{-3}$ \citep{Brocklehurst1971}; the \cite{Whitford1958}
reddening curve as parametrized for \cite{Miller1972} was used.

For \Hb, where absorption wings are clearly resolved, we fitted simultaneously
an absorption and an emission component, in order to obtain precise
measurements of the \Hb\ flux in emission; however, in \Ha, the absence of
absorption wings makes a reliable decomposition impossible. 
Again, we assume that the absorption in \Ha\ is negligible, so that the
derived \chbeta\ are actually lower limits.

For comparison, we also tabulate the values obtained by setting
EW$(\Ha_\mathrm{abs})=2$ \AA, as in the output spectra from the synthesis
models run in Sect.~\ref{Sect:Discussion}, we measured an 
$\mathrm{EW}(\Ha_\mathrm{abs}$) of about 1.5--2 \AA. For this latter case we
see a modest increase in the extinction coefficient.

The reddening corrected line intensities in each region are listed in 
Table~\ref{Table:emissionlines}. The most relevant line ratios are shown in  
Table~\ref{Table:ratiolines}.

Th \Ha\ flux emitted by the SF regions is about 60\% of the total \Ha\ flux
within the PMAS field of view.

\begin{deluxetable}{cccccccccccccc}
\tabletypesize{footnotesize}
\tabletypesize{\scriptsize}
\rotate
\tablewidth{0pt}
\tablecaption{Reddening corrected line intensity ratios}
\tablehead{
\colhead{Line}            & \colhead{Ion}             & 
\multicolumn{2}{c}{Knot \#1} & \multicolumn{2}{c}{Knot \#2} & 
\multicolumn{2}{c}{Knot \#3} & \multicolumn{2}{c}{Knot \#4} & 
\multicolumn{2}{c}{Knot \#5} & \multicolumn{2}{c}{Knot \#6} \\
\colhead{$(\AA)$}         &                           & 
\colhead{$F_{\lambda}$}   & \colhead{$-W_{\lambda}$}  & 
\colhead{$F_{\lambda}$}   & \colhead{$-W_{\lambda}$}  & 
\colhead{$F_{\lambda}$}   & \colhead{$-W_{\lambda}$}  & 
\colhead{$F_{\lambda}$}   & \colhead{$-W_{\lambda}$}  & 
\colhead{$F_{\lambda}$}   & \colhead{$-W_{\lambda}$}  &
\colhead{$F_{\lambda}$}   & \colhead{$-W_{\lambda}$}  
}
\startdata
3727 & [\ion{O}{2}]        & $\phn5.06\pm1.18$  &  $   40.3\pm2.9$ 
                           & $\phn4.29\pm0.80$  &  $   46.8\pm2.6$ 
                           & $   13.13\pm2.61$  &  $   25.5\pm0.7$
                           & $\phn4.84\pm0.62$  &  $   52.3\pm1.8$	  
                           & $\phn5.03\pm0.68$  &  $   47.0\pm1.9$
                           & $\phn3.50\pm0.34$  &  $   86.1\pm1.2$ \\
4861 & \Hb\                & $\phn1.00\pm0.00$  &  $\phn5.4\pm0.7$   
                           & $\phn1.00\pm0.00$  &  $\phn7.9\pm0.8$ 
                           & $\phn1.00\pm0.00$  &  $\phn1.4\pm0.2$
                           & $\phn1.00\pm0.00$  &  $\phn9.1\pm0.5$
                           & $\phn1.00\pm0.00$  &  $\phn7.4\pm0.5$ 
                           & $\phn1.00\pm0.00$  &  $   18.6\pm0.3$ \\
4959 & [\ion{O}{3}]        & $\phn0.89\pm0.15$  &  $\phn5.2\pm0.4$   
                           & $\phn0.48\pm0.07$  &  $\phn3.9\pm0.4$ 
                           & $\phn2.51\pm0.34$  &  $\phn3.8\pm0.1$
                           & $\phn0.61\pm0.05$  &  $\phn5.9\pm0.3$
                           & $\phn0.74\pm0.07$  &  $\phn6.0\pm0.4$ 
                           & $\phn0.79\pm0.03$  &  $   14.3\pm0.2$ \\
5007 & [\ion{O}{3}]        & $\phn2.13\pm0.30$  &  $   12.6\pm0.4$   
                           & $\phn1.64\pm0.18$  &  $   13.7\pm0.4$ 
                           & $\phn5.68\pm0.76$  &  $\phn8.9\pm0.1$ 
                           & $\phn1.89\pm0.12$  &  $   18.6\pm0.4$
                           & $\phn1.98\pm0.14$  &  $   16.3\pm0.4$ 
                           & $\phn2.36\pm0.07$  &  $   43.4\pm0.4$ \\
6300 & [\ion{O}{1}]        & $\phn0.20\pm0.06$  &  $\phn1.3\pm0.3$   
                           & $\phn0.16\pm0.03$  &  $\phn1.5\pm0.2$
                           & $\phn0.44\pm0.09$  &  $\phn1.0\pm0.1$ 
                           & $\phn0.18\pm0.03$  &  $\phn2.1\pm0.3$ 
                           & $\phn0.17\pm0.03$  &  $\phn1.7\pm0.3$
                           & $\phn0.10\pm0.01$  &  $\phn2.0\pm0.3$ \\
6548 & [\ion{N}{2}]        & $\phn0.25\pm0.07$  &  $\phn1.7\pm0.3$   
                           & $\phn0.25\pm0.05$  &  $\phn2.5\pm0.3$
                           & $\phn0.40\pm0.09$  &  $\phn1.0\pm0.1$  
                           & $\phn0.27\pm0.03$  &  $\phn3.3\pm0.3$  
                           & $\phn0.25\pm0.03$  &  $\phn2.6\pm0.2$ 
                           & $\phn0.21\pm0.01$  &  $\phn4.5\pm0.2$ \\
6563 & \Ha\                & $\phn2.86\pm0.55$  &  $   19.7\pm0.4$   
                           & $\phn2.86\pm0.43$  &  $   28.9\pm0.6$
                           & $\phn2.86\pm0.52$  &  $\phn7.5\pm0.1$ 
                           & $\phn2.86\pm0.24$  &  $   35.6\pm0.6$ 
                           & $\phn2.86\pm0.27$  &  $   29.9\pm0.4$ 
                           & $\phn2.86\pm0.12$  &  $   60.1\pm0.3$ \\
6584 & [\ion{N}{2}]        & $\phn0.58\pm0.13$  &  $\phn4.0\pm0.3$   
                           & $\phn0.62\pm0.10$  &  $\phn6.2\pm0.4$ 
                           & $\phn1.19\pm0.22$  &  $\phn3.1\pm0.1$
                           & $\phn0.67\pm0.06$  &  $\phn8.3\pm0.3$
                           & $\phn0.62\pm0.06$  &  $\phn6.4\pm0.3$
                           & $\phn0.53\pm0.02$  &  $\phn6.4\pm0.3$ \\
6717 & [\ion{S}{2}]        & $\phn0.58\pm0.13$  &  $\phn4.2\pm0.4$    
                           & $\phn0.49\pm0.08$  &  $\phn5.1\pm0.3$
                           & $\phn0.89\pm0.17$  &  $\phn2.5\pm0.1$ 
                           & $\phn0.67\pm0.06$  &  $\phn8.8\pm0.3$ 
                           & $\phn0.57\pm0.06$  &  $\phn6.2\pm0.4$ 
                           & $\phn0.40\pm0.02$  &  $\phn0.0\pm0.4$ \\
6731 & [\ion{S}{2}]        & $\phn0.44\pm0.10$  &  $\phn3.2\pm0.3$    
                           & $\phn0.35\pm0.06$  &  $\phn3.7\pm0.3$
                           & $\phn0.86\pm0.16$  &  $\phn2.4\pm0.1$
                           & $\phn0.51\pm0.05$  &  $\phn6.7\pm0.4$
                           & $\phn0.39\pm0.05$  &  $\phn4.3\pm0.3$
                           & $\phn0.30\pm0.02$  &  $\phn0.0\pm0.3$ \\[6pt]
\chbeta   &         &   
\multicolumn{2}{c}{$0.23\pm0.19$}     & \multicolumn{2}{c}{$0.22\pm0.14$}  &
\multicolumn{2}{c}{$\phn0.91\pm0.18$} & \multicolumn{2}{c}{$0.21\pm0.08$}  &
\multicolumn{2}{c}{$0.27\pm0.09$}     & \multicolumn{2}{c}{$\phn0.21\pm0.04$} \\
\chbeta$^{*}$                         & 				   & 
\multicolumn{2}{c}{$0.35\pm0.19$}     & \multicolumn{2}{c}{$0.30\pm0.14$}  &
\multicolumn{2}{c}{$\phn1.23\pm0.18$} & \multicolumn{2}{c}{$0.27\pm0.08$}  &
\multicolumn{2}{c}{$0.35\pm0.09$}     & \multicolumn{2}{c}{$\phn0.22\pm0.04$} \\
$F(\Hb)$                              & 				   & 
\multicolumn{2}{c}{$1.74\pm0.84$}     & \multicolumn{2}{c}{$2.51\pm0.94$}  &
\multicolumn{2}{c}{$18.27\pm8.91$}    & \multicolumn{2}{c}{$4.09\pm0.86$}  &
\multicolumn{2}{c}{$3.63\pm0.87$}     & \multicolumn{2}{c}{$17.27\pm1.74$} \\	   
\enddata
\tablecomments{Reddening-corrected line intensities, normalized to $F(\Hb)=1$, 
for the individual regions selected in Mrk~409.
The reddening coefficient, \chbeta, and the corrected \Hb\ flux, 
$F(\Hb)$ ($\times 10^{-15}$ \ergscms), were computed assuming no
underlying \Ha\ absorption.
For comparison, we also list \chbeta$^{*}$, the reddening coefficient derived 
by assuming EW$(\Ha_{\mathrm abs})=2$ \AA.
\label{Table:emissionlines} }
\end{deluxetable}

\subsubsection{Constraining the Ionization Mechanism}
\label{SubSubSect:IonizationMechanisms}

The ionization mechanisms acting in a galaxy and their spatial distribution
can be studied by means of the line ratio diagrams for the usual diagnostic
lines \citep{Baldwin1981,Veilleux1987}. These diagnostic diagrams help 
distinguish among the different ionization mechanisms (photoionization from
young stars, photoionization by a power law continuum source and shock-wave
heating) operating in a galaxy.

\begin{figure}   
\begin{center}
\includegraphics[angle=-90,width=\textwidth]{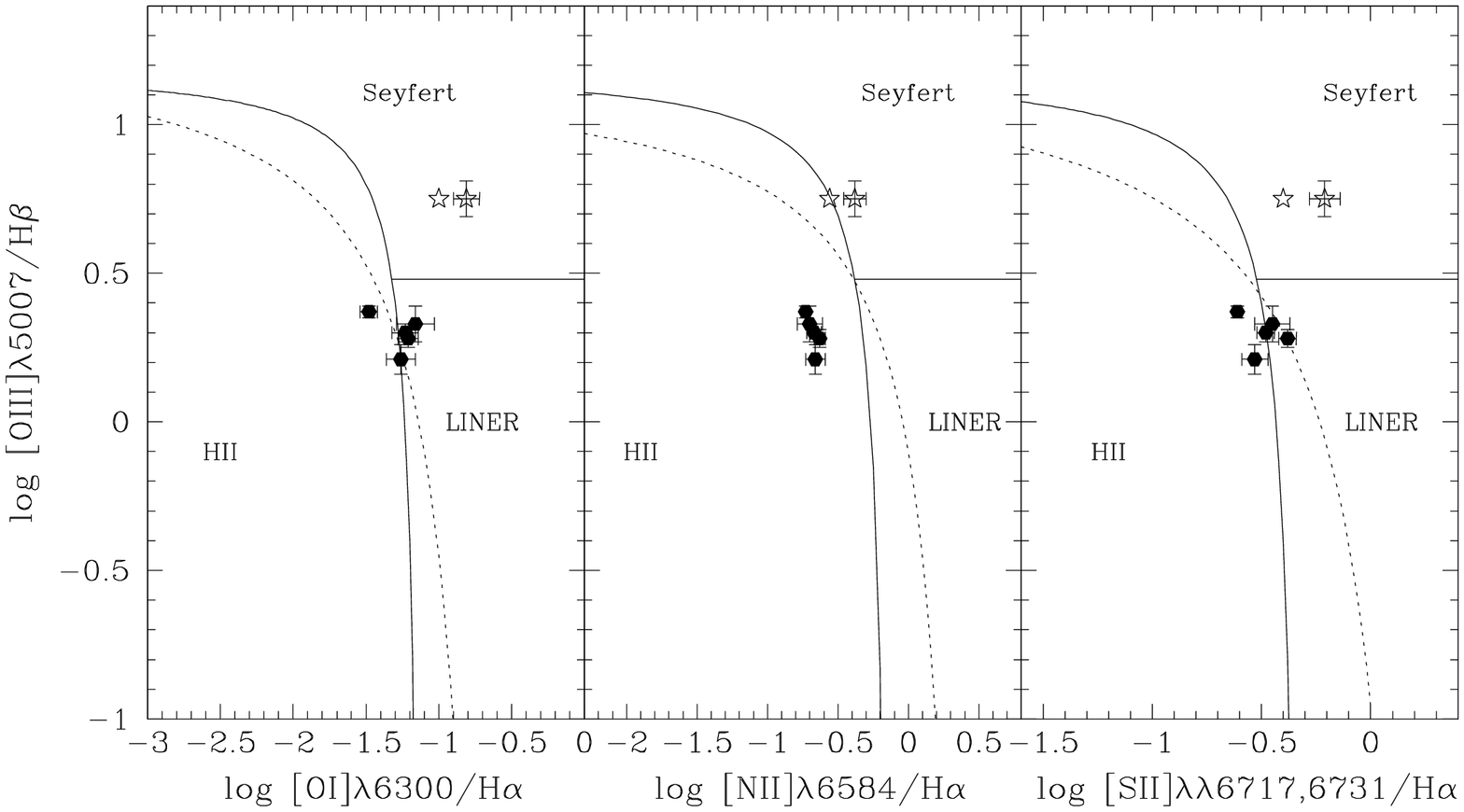}
\caption{Optical  emission-line  diagnostic  diagrams, separating  Seyfert
galaxies and LINERs from \ion{H}{2} region-like objects. Solid lines are the
empirical  borders  from  \cite{Veilleux1987},  while  dotted  lines  are  the
theoretical borders from \cite{Kewley2001}.  The line ratios for the nuclear SF
region  of Mrk~409 (\#3) and the five  circumnuclear SF  regions  (\#1--2 and
\#4--6) are plotted with open  and filled  symbols, respectively. For the
nuclear region, we also mark (open star without  errorbar) the  position
corresponding  to assuming a correction for underlying  \Ha\  absorption of
EW$(\Ha_\mathrm{abs})=4$  \AA\ (for all other knots,  this  shift would  be
unnoticeable). Note that, in all three classification diagrams, the nuclear
SF region falls in the Seyfert regime.}
\label{Fig:diagnostic}
\end{center}
\end{figure}

The position of the six regions under study on the diagnostic diagram
[\ion{O}{3}]~$\lambda5007$/\Hb\ vs [\ion{O}{1}]~$\lambda6300$/\Ha,
[\ion{N}{2}]~$\lambda6584$/\Ha\ and
[\ion{S}{2}]~$\lambda\lambda6717,\;6731$/\Ha\ is shown in
Figure~\ref{Fig:diagnostic}. The empirical boundaries between the different
zones (from \citealp{Veilleux1987}), as well as the theoretical boundaries
proposed by \cite{Kewley2001}, are also plotted.

\ion{H}{2}-like  ionization (i.e. star  formation) seems  to be  the dominant
mechanism   in  the  five SF knots  conforming the ring  (see also
Table~\ref{Table:ratiolines}  for the  specific values  of the  emission line
ratios). The line  ratios of the nuclear region  \#3, however, are consistent
with those of Seyfert galaxies. We  came to the same  conclusion assuming an
extreme correction for the underlying \Ha\ absorption of
EW$(\Ha_\mathrm{abs})=4$ \AA\ for the  nuclear  region. The  possibility of  the
nuclear  region of  Mrk~409 containing  a non-thermal  energy source  cannot be
discarded. Indirect  evidence for this interpretation comes  from the failure
of purely  stellar synthetic  spectra to reproduce  the observed  spectrum of
that region (see Sect.~\ref{Sect:Discussion}).

\subsubsection{Physical parameters and chemical abundances}
\label{SubSubSect:ChemicalAbundances}

The physical properties and oxygen abundances for Mrk~409 are listed in
Table~\ref{Table:ratiolines}. 

The electron density (N$_{e}$) for the individual SF knots was derived from
the [\ion{S}{2}]~$\lambda\lambda6717,\;6731$ \AA\ line ratio using the task 
\textsc{temden} in the IRAF \textsc{nebular} package 
\citep{deRobertis1987,ShawDufour1995}. 

The "direct" oxygen abundance method relies on measurements of
temperature-sensitive auroral lines (usually [\ion{O}{3}]~$\lambda4363$), 
which are weak and often not detected in oxygen-rich systems (above $\sim
12+\log(\mathrm{O/H}) = 8.3-8.4$). This is the case of the spectra of
Mrk~409,  in which either we fail to detect the [\ion{O}{3}]~$\lambda4363$ 
emission line, or, in the cases where we do detect the
[\ion{O}{3}]~$\lambda4363$ line, the S/N is too low for a reliable flux
determination. 

Different empirical and theoretical metallicity calibrations have been devised
over the past two decades as an alternative for estimating oxygen abundances
without any direct electron temperature measurements
\citep{Pagel1979,Alloin1979,Edmunds1984,McGaugh1991,Vilchez1996,
Pettini2004,Pilyugin2006}.
These calibrations are referred to as strong-line methods since they are based
on ratios built on strong emission lines ([\ion{O}{2}]~$\lambda3727$,
[\ion{O}{3}]~$\lambda\lambda4959,\;5007$,
[\ion{N}{2}]~$\lambda6584$,[\ion{S}{2}]~$\lambda\lambda6717,\;6731$,
[\ion{S}{3}]~$\lambda\lambda9069,\;9532$). We have used the strong-line
methods from \cite{McGaugh1991}[hereafter M91] and \cite{Pettini2004}[hereafter
PP04] to derive the oxygen abundances of all selected SF knots, except the
nuclear region \#3 (see below).  

Efforts have been made to refine the calibration of the $R_{23}$ parameter.
One of the most successful is the calibration by M91, which is based on
photoionization models --- the analytic expressions for the M91 theoretical
models can be found in \cite{Kobulnicky1999}. This calibration can be as precise
as 0.2 dex \citep{Kobulnicky1999}. However, due to the well-known double-valued
behavior between $R_{23}$ and O/H, this method requires some prior knowledge
of the galaxy's metallicity, in order to place it on the correct branch of the
curve. We used the ratio [\ion{N}{2}]~$\lambda$6584/\Ha\ to break the R$_{23}$
degeneracy; in all the SF regions, log([\ion{N}{2}]~$\lambda6584$/\Ha) is $>
-1.1$, indicating that the knots lie on the upper branch --- the division
between the $R_{23}$ upper and lower branches takes place between  $-1.3 <
\log($[\ion{N}{2}]~$\lambda6584/\Ha) < -1.1$ \citep{Kewley2008}.

On the other hand, PP04 revised the N2 (N2 $\equiv$
log{[\ion{N}{2}]~$\lambda6584$/\Ha}) and O3N2 (O3N2 $\equiv$
log{([\ion{O}{3}]~$\lambda5007$/\Hb/[\ion{N}{2}]~$\lambda$6584/\Ha)}) indices,
using 137 extragalactic HII regions.  The uncertainties in metallicity
determinations based on these calibrations amount to $\sim 0.38$ and 0.25 dex,
respectively. 

Inspection of Fig.~\ref{Fig:diagnostic} shows knot \#3 to be located in the
Seyfert zone of the diagnostic diagram. For this region, we estimate therefore
the metallicity using the chemical abundance calibrations proposed by
\cite{StorchiBergmann1998}. These calibrations were derived for the range $8.4
\leq 12 + \log(\mathrm{O/H}) \leq 9.4$ and tested using a sample of Seyfert
galaxies and LINERs that have \ion{H}{2} regions in the vicinity of their
nucleus.

\begin{deluxetable}{lccccccc}
\tabletypesize{\footnotesize}
\tabletypesize{\scriptsize}
\rotate
\tablewidth{0pt}
\tablecaption{Line ratios, physical parameters and chemical abundances}
\tablehead{
\colhead{Parameter} & 
\colhead{{Knot \#1}} & 
\colhead{{Knot \#2}} & 
\colhead{{Knot \#3}} & 
\colhead{{Knot \#4}} & 
\colhead{{Knot \#5}} &  
\colhead{{Knot \#6}} 
}
\startdata
$\log(\frac{[\mbox{\ion{O}{3}}]\;\lambda5007}{\Hb})$                & 
       \phs$0.33\pm0.06$ & \phs$0.22\pm0.05$ & \phs$0.75\pm0.06$    & 
       \phs$0.28\pm0.03$ & \phs$0.30\pm0.03$ & \phs$0.37\pm0.02$    \\
$\log(\frac{[\mbox{\ion{N}{2}}]\;\lambda6584}{\Ha})$                & 
       $   -0.70\pm0.09$ & $   -0.66\pm0.07$ & $   -0.38\pm0.08$    & 
       $   -0.63\pm0.04$ & $   -0.67\pm0.05$ & $   -0.73\pm0.03$    \\
$\log(\frac{[\mbox{\ion{S}{2}}]\;\lambda\lambda6717,\;6731}{\Ha})$  & 
       $   -0.45\pm0.08$ & $   -0.53\pm0.06$ & $   -0.21\pm0.07$    & 
       $   -0.38\pm0.04$ & $   -0.48\pm0.04$ & $   -0.61\pm0.02$    \\ 
$\log(\frac{[\mbox{\ion{O}{1}}]\;\lambda6300}{\Ha})$                & 
       $   -1.16\pm0.13$ & $   -1.26\pm0.10$ & $   -0.81\pm0.09$    & 
       $   -1.21\pm0.07$ & $   -1.23\pm0.09$ & $   -1.46\pm0.06$    \\[2pt] 
\hline
$N_\mathrm{e}$                 & 117       & {\bf $\leq 100$}  & 516	  &  102     & {\bf $\leq 100$} & {\bf $\leq 100$}  \\
$12 +\log(\mathrm{O/H})^{a}$   & 8.40      & 8.45              & \nodata   &  8.44    &  \nodata             & 8.38              \\
$12 +\log(\mathrm{O/H})^{b}$   & 8.46      & 8.49              & \nodata   &  8.51    & 8.48             & 8.43              \\
$12 +\log(\mathrm{O/H})^{c}$   & 8.40      & 8.53              & \nodata   &  8.45    & 8.42             & 8.54              \\
$12 +\log(\mathrm{O/H})^{d}$   & \nodata   & \nodata           & 8.44      & \nodata  & \nodata	        & \nodata           \\
$\log(\mathrm{N/O})^{e}$       & $-1.13$   & $-1.02$           & $-1.28$   & $-1.05$  &$-1.11$           & $-1.01$           \\
\enddata
\tablecomments{The line ratios have been corrected for galactic extinction. 
The quoted uncertainties include both measurement errors and the uncertainty
in the calibration factor. 
(a) Oxygen abundances computed using the fit to the relationship between 
$T_\mathrm{e}$ metallicities and the ([\ion{O}{3}]/\Hb)/([\ion{N}{2}]/\Ha) in 
\cite{Pettini2004}; 
(b) oxygen abundances obtained from the relationship between $T_\mathrm{e}$ 
metallicities and the [\ion{N}{2}]/\Ha\ ratio given by \cite{Pettini2004}; 
(c) oxygen abundances derived using the \cite{McGaugh1991} calibration for the 
upper branch; 
(d) oxygen abundance computed from chemical abundance calibrations for 
Seyfert galaxies \citep{StorchiBergmann1998};
(e) [\ion{N}{2}]/[\ion{O}{2}] was used to obtain the N/O abundance ratio 
\citep{PerezMontero2005}.}
\label{Table:ratiolines}
\end{deluxetable}
\clearpage

\subsection{Kinematics of the ionized gas}
\label{SubSect:Kinematics}

\begin{figure*}[h]
\mbox{
\centerline{
\hspace*{-0.5cm}\subfigure{\includegraphics[width=7.0cm]{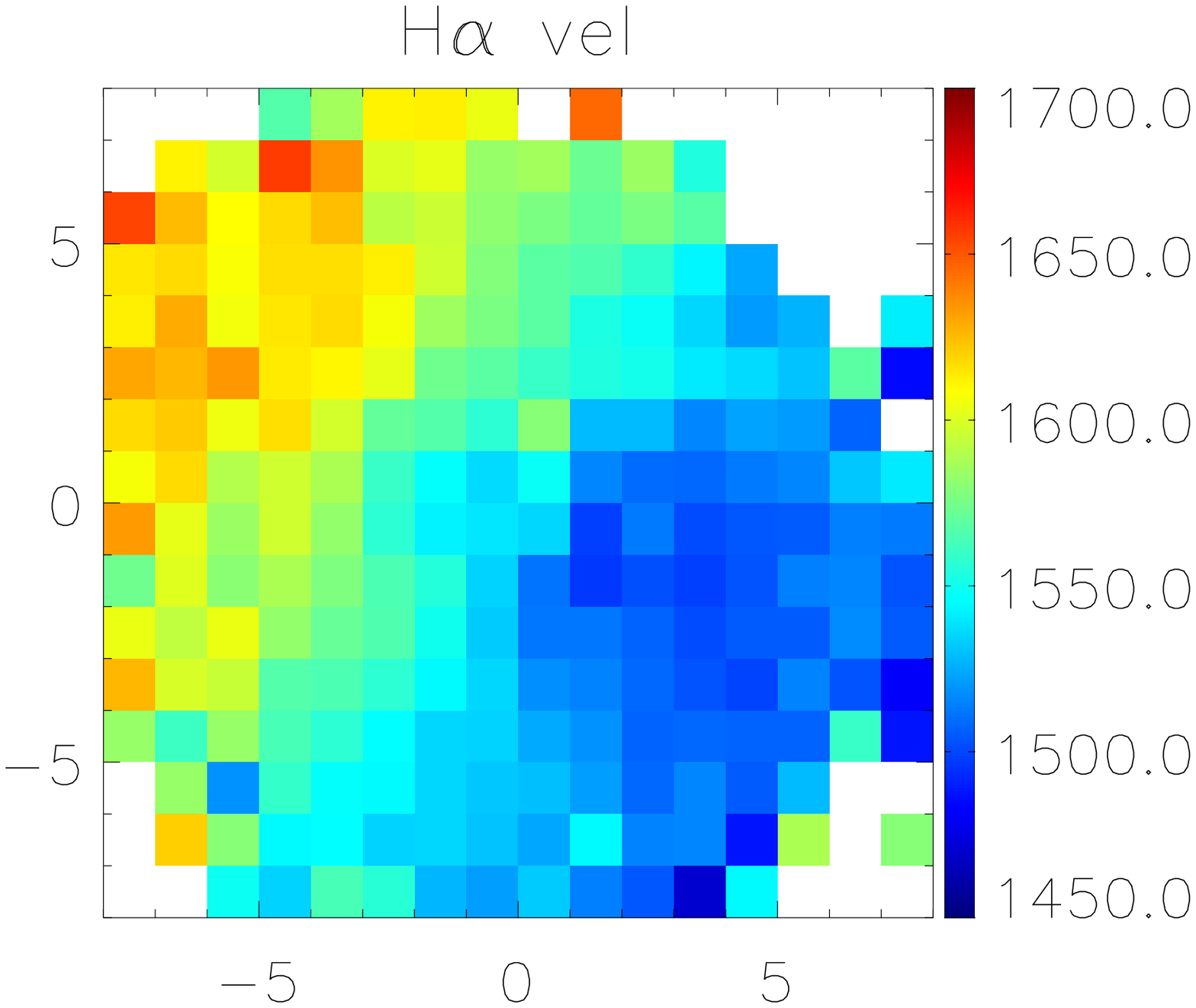}}
\hspace*{0.5cm}\subfigure{\includegraphics[width=7.0cm]{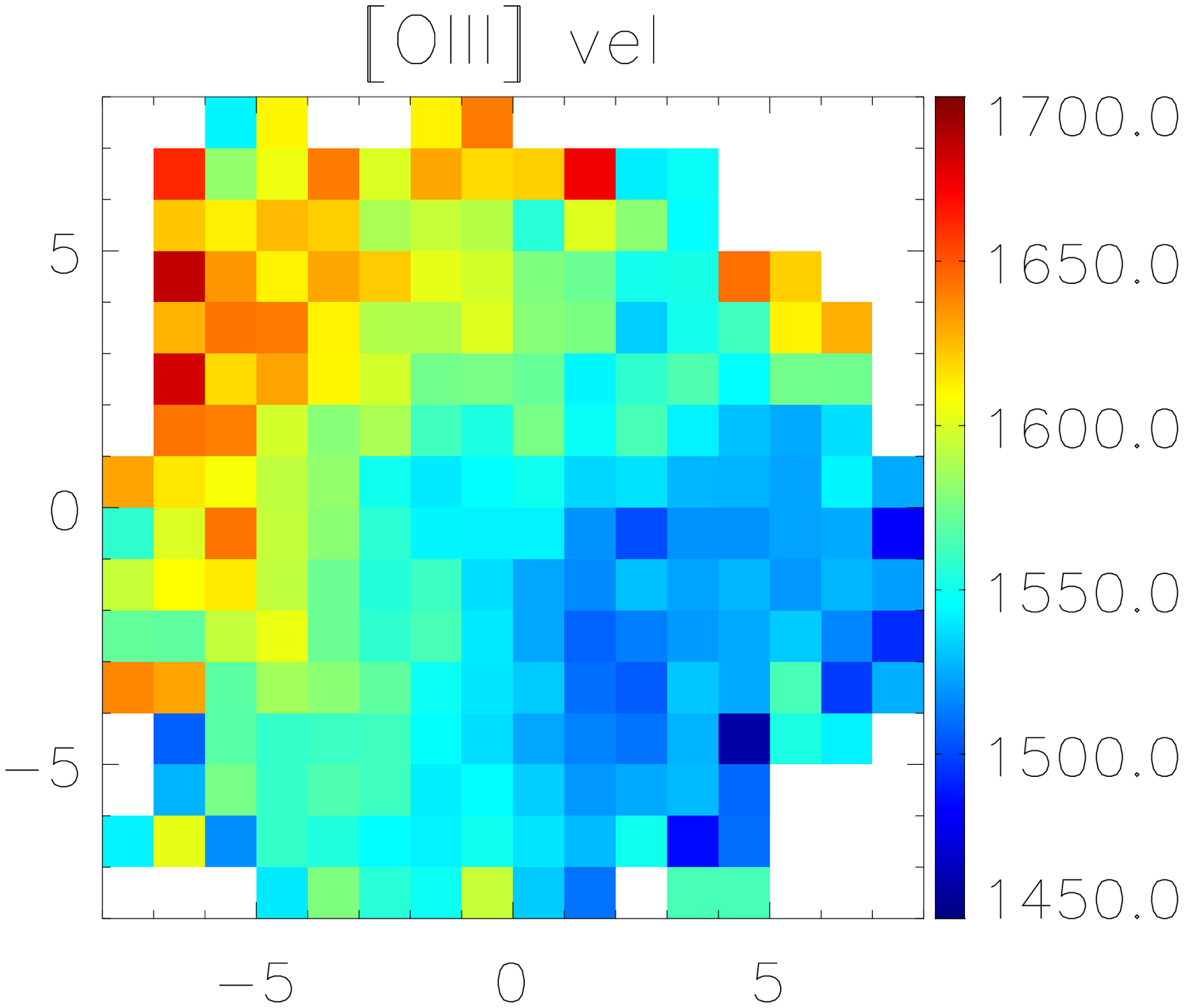}}
}}   
\caption{Velocity field of the ionized gas in the central region of
Mrk~409 in the [\ion{O}{3}] and \Ha\ lines. Axis units are arcseconds; 
north is up, east to the left.}
\label{Fig:velocity}
\end{figure*}

Figure~\ref{Fig:velocity} shows the velocity field of the ionized gas, derived
from the [\ion{O}{3}]~$\lambda5007$ and the \Ha\ emission lines.  Both
velocity maps show essentially the same regular pattern, although  in the
[\ion{O}{3}]~$\lambda5007$ map it appears slightly more distorted; the galaxy
displays an overall smooth rotation pattern along a northwest-southeast axis.

\begin{figure*}[h]
\mbox{
\centerline{
\hspace*{0.0cm}\subfigure{\includegraphics[width=0.5\textwidth]{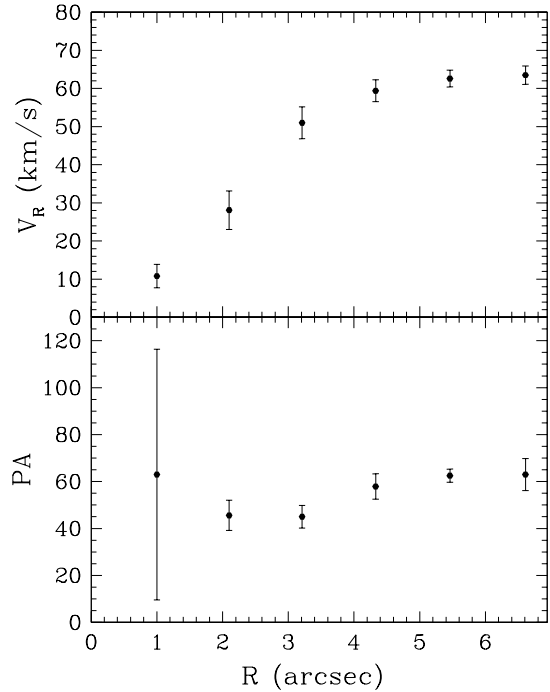}}
}}   
\caption{Velocity curve and position angle profile of the kinematical major 
axis (measured from north through east) as determined by a tilted-ring model 
analysis.}
\label{Fig:velocityprofile}
\end{figure*}

Figure~\ref{Fig:velocityprofile} shows the velocity profile and the position
angle of the kinematical axis, as determined by a tilted-ring analysis of the
\Ha\ velocity field based on the \textit{kinemetry} method developed by
\cite{Krajnovic2006}.

The velocity rises rapidly in the inner 4 arcsecs, then flattens out at about
60 \kmsec. Adopting for simplicity the virial formula valid for spherical
symmetry, we can estimate the mass from the relation: $M(R)=2.32\times 10^5\,
R\, v_\mathrm{circ}^2(R)$ \citep{Lequeux1983}, where $R$ is in kpc,
$v_\mathrm{circ} = v_\mathrm{obs}/sin(i)$ is the circular velocity, $i$ is the
inclination angle, and $M$ is expressed in solar units.

The apparent flattening of Mrk~409 is $q=b/a\simeq 0.80$ (SDSS, Data Release
6), which, for an intrinsic flattening $q_\mathrm{intr} \lesssim 0.3$ gives
an inclination angle $i \simeq 38$.
Taking an observed rotational velocity $v\simeq 60$ \kmsec\ at the position 
of the SF ring, we obtain an indicative virial mass of 
$\simeq1.4 \times 10^9$ \msun\ interior to the SF ring 
($R_\mathrm{ring}\approx0.6$ kpc).

\section{Discussion}
\label{Sect:Discussion}

As it was already stated in the introduction, one of the main open questions in
BCD research is how the star formation originates and propagates across these
galaxies. The spatial distribution of the SF knots in the central part of
Mrk~409 --- a nuclear starburst surrounded by a SF ring ---  makes this
galaxy a paradigm case of study. 

An extended circumnuclear SF zone may play an important role on BCD evolution,
contributing to the build-up of the stellar component and to the chemical
enrichment of the Interstellar Medium (ISM). In addition, the possibility of
Mrk~409 hosting a non-thermal nuclear energy source (see
Sect.~\ref{SubSubSect:IonizationMechanisms}), makes this object even more   
intriguing.  In this line, \cite{Izotov2008} reported recently on the
detection of AGN candidates in four metal-poor dwarf galaxies, and 
circumnuclear SF activity is known to exist in several AGNs hosted by normal
galaxies \citep{Wilson1986,Kotilainen2000,Gonzalez-Delgado2001}.

GdP03 suggested that the SF ring in Mrk~409 could have been formed as the result
of the interaction of a starburst-driven shock with the sourrounding
ISM.  These authors performed echelle spectroscopy of the
central part of Mrk~409 and found a doubly-peaked \Ha\ emission line profile,
with its two peaks shifted by $-54\pm3$ and $37\pm3$ \kmsec\ relative to the
systemic velocity of the stars (presumably because of the lower spectral
resolution, we do not see this feature in our spectra). But their data,
covering only a thin slice of the BCD's central part, did not allow to firmly
validate this scenario.  

This idea appears also viable from the viewpoint of observational and
theoretical work that indicates that BCDs  experience multiple bursts of star
formation in their central part through their lifetimes (e.g.
\citealp{Krueger1994,MasHesseKunth1999}). The mechanical energy released by
the collective action of stellar winds from massive stars and SNe can readily
result in large-scale expansive motions in the ISM reflected on e.g. \Ha\
super-shells out to kpc scales from the starburst region
\citep{Meurer1992,Marlowe1995,Cairos2001}. Observations of BCDs indicate
expansive velocities between a few 10 \kmsec\  \citep[see
e.g.][]{Meurer1992,Marlowe1995,Martin1997,vanEymeren2007} and  $\ga 200$
\kmsec\ (He~2-10, \citealp{PapaderosFricke1999,Johnson2000}; Haro~2,
\citealp{Legrand1997}).

An alternative hypothesis, which must also be taken into account, is that the
ring found in Mrk~409 is a resonance ring (RR), similar to those observed in
barred galaxies  \citep{Buta1988,ButaPurcell1998,Kotilainen2000}. RRs are
thought to result from efficient gas inflow onto the central region of a
galaxy, driven by the bar \citep{CombesGerin1985}, and its collapse at the
radius of the inner Lindblad resonance. While RRs are typically observed in
galaxies of normal mass, they may as well exist in lower-mass systems, such as
Mrk~409. Note that the gas-phase metallicity in the SF ring in Mrk~409 is in
the range of values determined in circumnuclear SF regions in normal galaxies
\citep{Diaz2007}, and also that the SF ring compares well in size (R$\sim$600
pc) with RRs in normal  galaxies --  the average radius of galactic RRs is 750
pc \citep{ButaCrocker1993} with values as small as $\sim$200 pc
\citep{Comeron2009}.

In order to distinguish between the two aforementioned scenarios it is
important to study the SFH of the individual knots along the ring. In the
framework of the RR hypothesis, it is reasonable to assume that this feature
has been stable over several orbital periods (a few $10^8$ yr), forming stars
at a nearly constant rate, whereas the blast wave hypothesis, on the other
hand, requires that the SF ring as a whole was created nearly instantaneously
in the recent past.

Our PMAS data, mapping the nucleus and the SF ring, permit, despite their
moderate spectral resolution, the investigation of this issue. In order to
constrain the SFH of each SF knot, we interpret their integrated spectrum 
(Section \ref{SubSect:IntegratedSpectroscopy}) by means of spectral synthesis
models. To check the consistency of our results, we use both the population
synthesis code {\sc starlight} 
\citep{CidFernandes2004,CidFernandes2005a,CidFernandes2005b,GarciaRissmann2005}
and the evolutionary spectral synthesis code {\sc pegase} \citep{Fioc1997}.

We use {\sc starlight} to synthesize the observed stellar continuum in each SF
region as due to the superposition of single-age stellar populations (SSPs) of
different ages and metallicities. The SSP library used is based on stellar
models by \cite{Bruzual2003} for a Salpeter \textit{initial mass function}
(IMF), five metallicities (between $Z_{\odot}/50$ and 2.5$\,Z_{\odot}$) and 59
ages (between 1 Myr and 13 Gyr). Prior to modelling, the observed spectra were
corrected for interstellar extinction (using the \chbeta\ derived without 
stellar absorption correction in \Ha), and spectral regions containing strong
emission lines or skyline subtraction residuals were flagged out.

Fig.~\ref{Fig:Starlight} (top panel) displays the
extinction-corrected spectra of the circumnuclear SF regions \#1, \#2 and
\#4--6, normalized at 4200 \AA\ (orange color)  together with the
best-fitting synthetic spectra overlaid in black.  Flagged intervals in each
spectrum are marked with shaded vertical bars. The smaller panels to the
upper-right and lower-right show the age distribution of the SSPs selected by
{\sc starlight} as a function of, respectively, their luminosity and mass
fraction in \%. We find that the spectra of
all circumnuclear regions are well fit by the superposition of a several  Gyr
old stellar component with a young ($<$50 Myr) population which  dominates the
optical light. In none of the regions do {\sc starlight} models indicate an
intermediate-age  stellar population bridging the gap between 0.1 and $\sim$1
Gyr. All solutions imply no or little ($A_V\leq0.14$ mag) intrinsic extinction
in excess to that inferred from the measured \chbeta. According to the fits, the
formation of the underlying host galaxy of Mrk~409  was accomplished between
$\sim 1$ and $\sim 10$ Gyr ago, and at least 3/4 of its stellar mass was at
place $\sim 6$ Gyr ago.  This is in good agreement with the red \br\ color of
$\sim1.1$ mag  of the host of Mrk~409 (GdP03).

\begin{figure}   
\centerline{  
\hspace*{0.0cm}\includegraphics[angle=-90.0,viewport=180 40 520 730,
clip,width=0.95\textwidth]{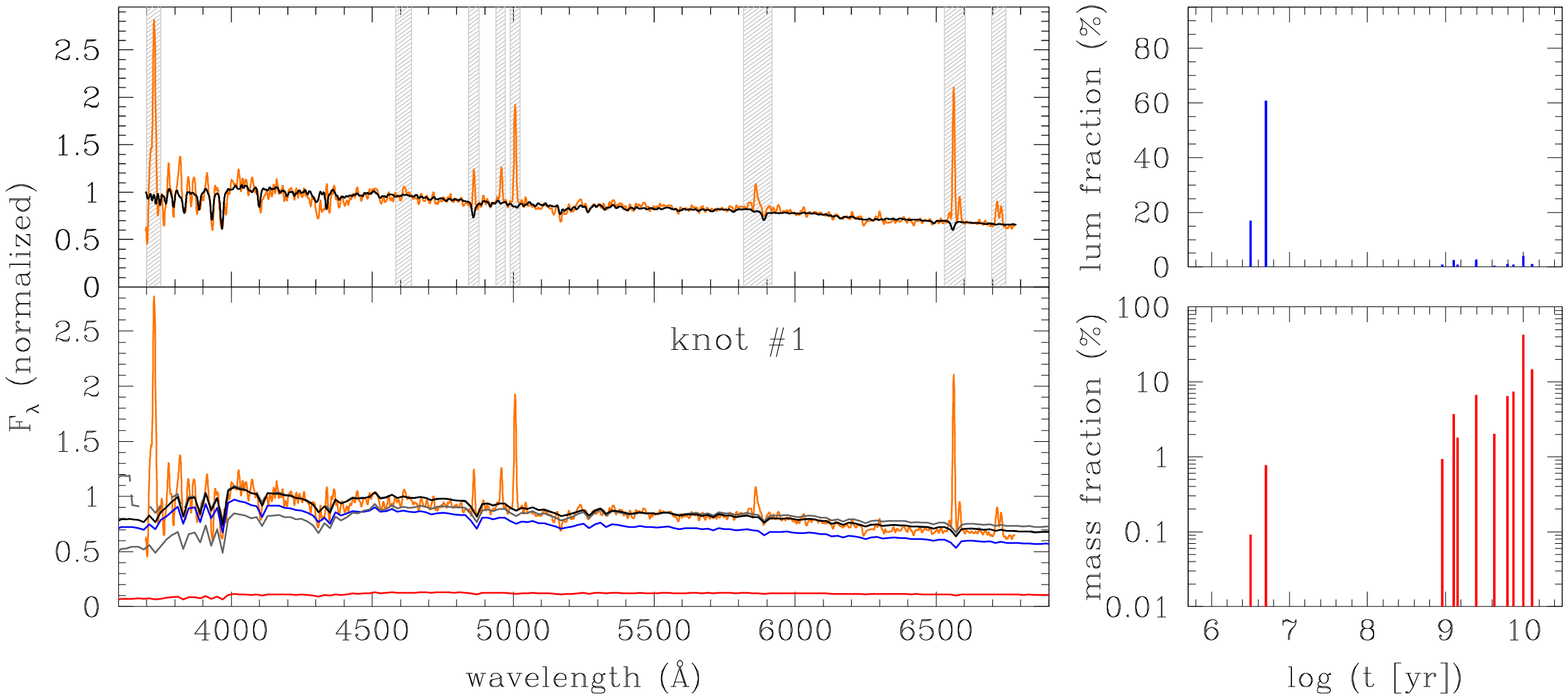}
}
\centerline{ 
\hspace*{0.0cm}\includegraphics[angle=-90.0,viewport=180 40 520 730,
clip,width=0.95\textwidth]{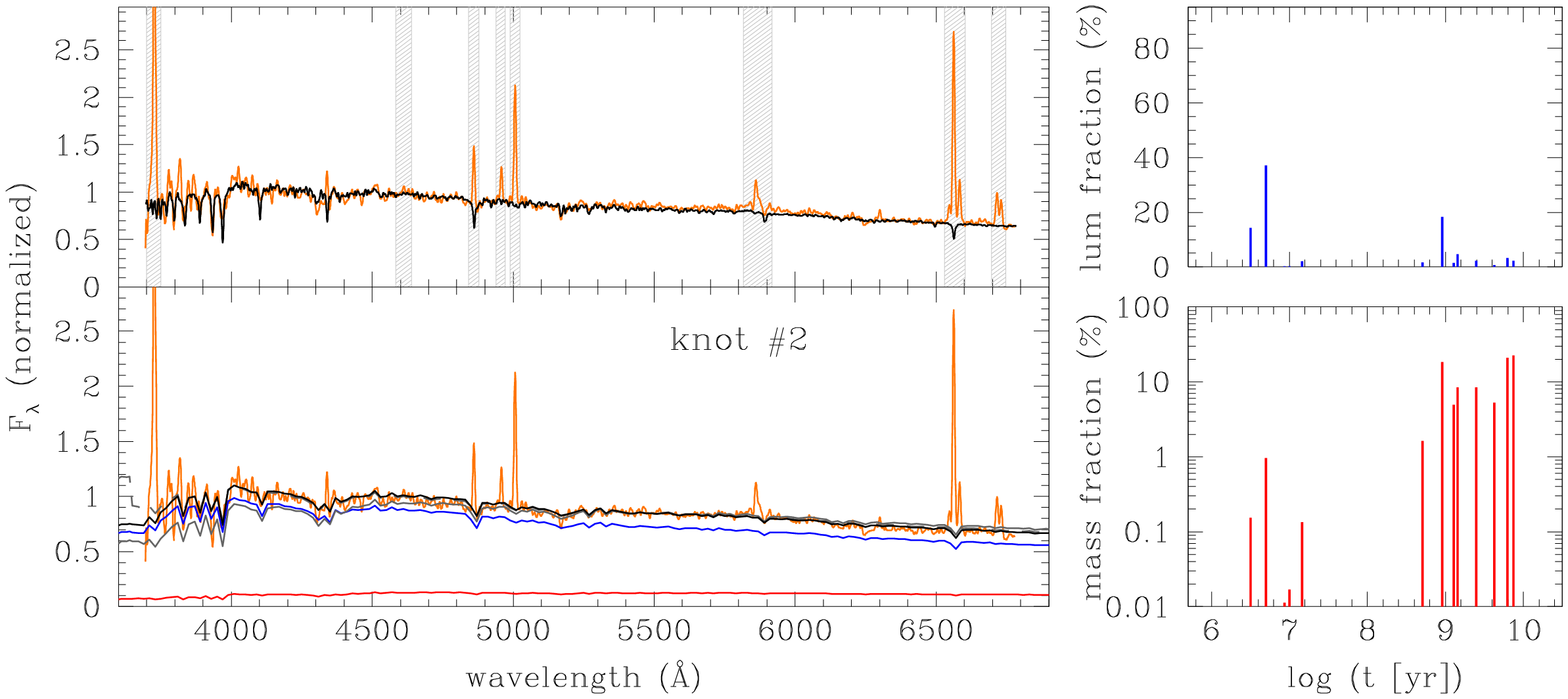}
}
\caption{Spectra of the five selected regions in the circumnuclear
star-forming ring in Mrk~409, corrected for extinction and normalized by the 
observed flux at 4200 \AA\ (orange color).
The spectra are arranged from region \#1 (top) to region \#6 (bottom).
The best-fitting synthetic spectral energy distribution
(SED) for the stellar continuum, calculated with the population synthesis code 
STARLIGHT \citep{CidFernandes2005a}, are shown in the upper panel in black. 
Flagged intervals are marked with shaded vertical bars.
The smaller upper-right and lower-right panels show the age 
distribution of the stellar populations in the synthetic SEDs
as a function of, respectively, their luminosity (at 4200 \AA) and mass 
contribution in \%.
 The SEDs corresponding to the best-fitting Pegase~2.0 models \texttt{inst}, 
\texttt{c0} and \texttt{c1} are shown in the lower panel with the black, 
solid-gray and dashed-gray curve, respectively. 
The blue and red curves illustrate the young and old stellar component, 
as derived from the \texttt{inst} model.
}
\label{Fig:Starlight}
\end{figure}

\begin{figure}  
\figurenum{8} 
\centerline{
\hspace*{0.0cm}\includegraphics[angle=-90.0,viewport=180 40 520 730,
clip,width=0.95\textwidth]{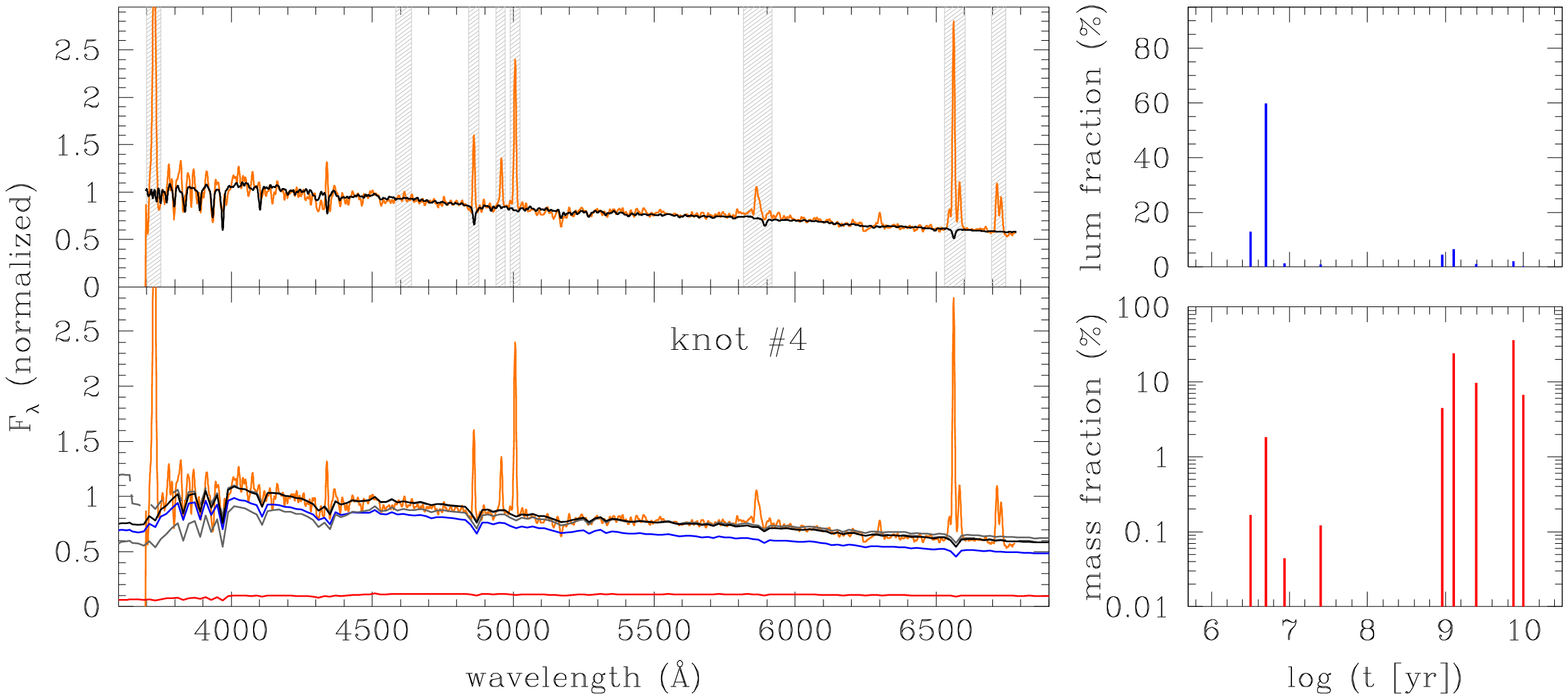}
}
\centerline{  
\hspace*{0.0cm}\includegraphics[angle=-90.0,viewport=180 40 520 730,
clip,width=0.95\textwidth]{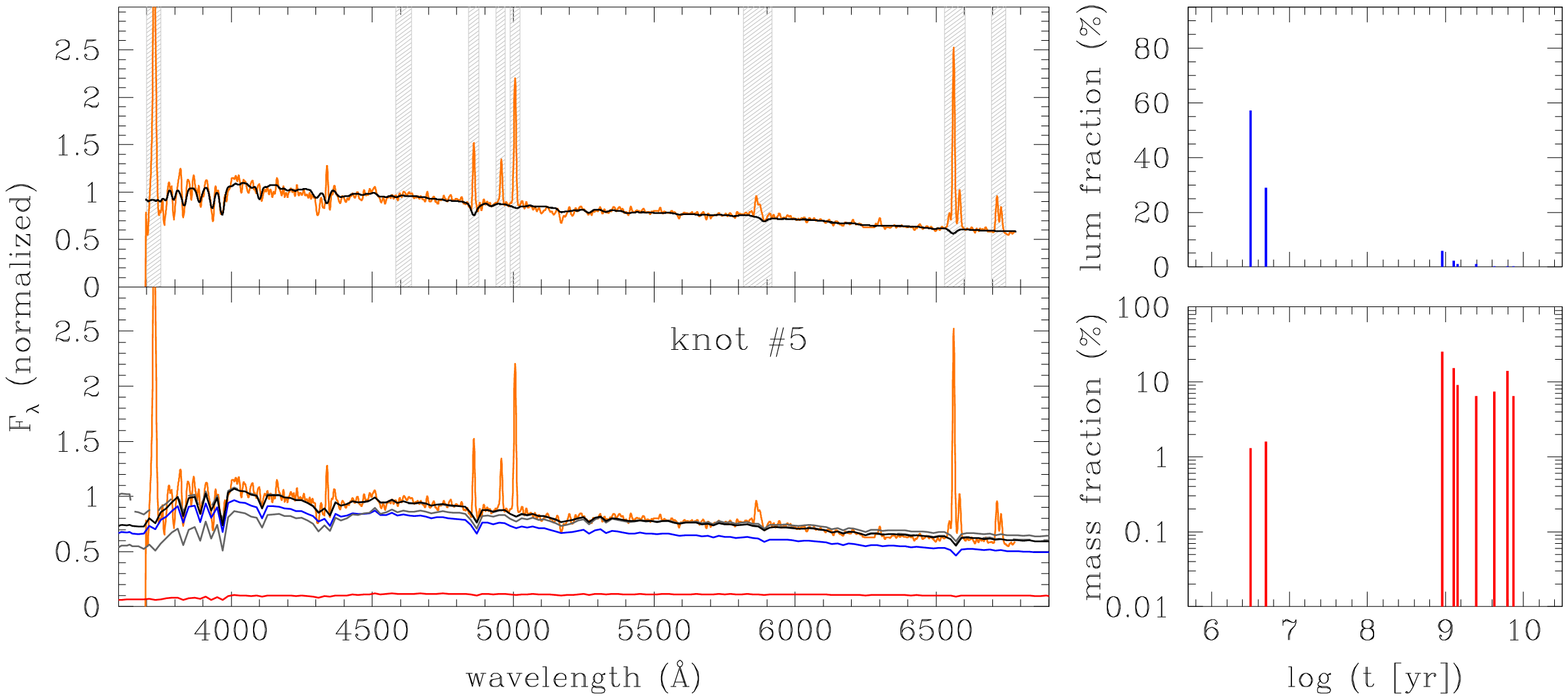}
}
\centerline{  
\hspace*{0.0cm}\includegraphics[angle=-90.0,viewport=180 40 520 730,
clip,width=0.95\textwidth]{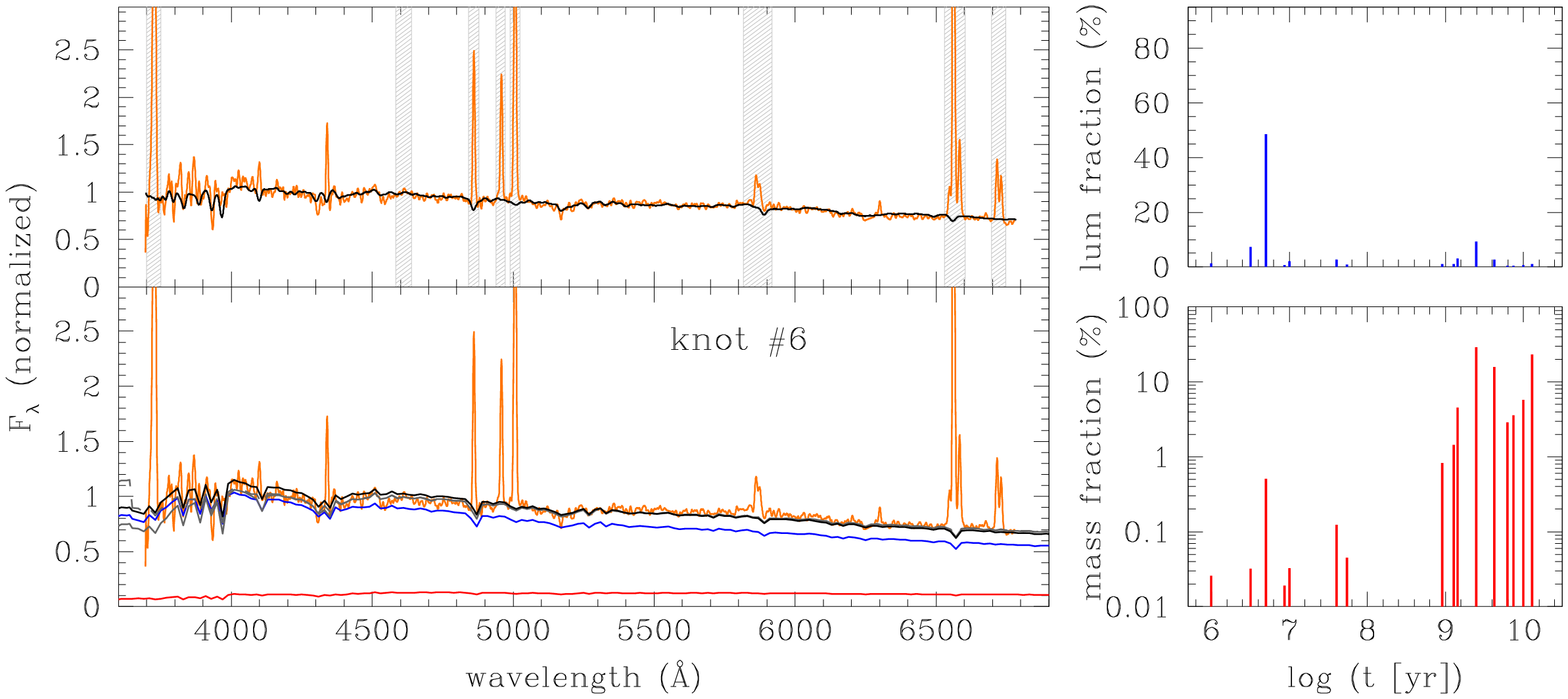}
}
\label{Fig:Starlight_2}
\caption{Continued.}
\end{figure}

Secondly, we apply a modified version of  {\sc pegase 2.0} to study whether
the SFH along the ring is better approximated by an instantaneous or a
continuous  model (models \texttt{inst} and \texttt{c0}--\texttt{c2}, 
respectively, in Table \ref{tab:evol}). In this approach, each of the observed
spectra is modeled as due to the superposition of the SED of an old and a
young stellar population describing, respectively, the underlying old host
galaxy and the SF ring. We use the same input spectra as for {\sc starlight}
after resampling to the  lower spectral resolution of {\sc pegase}. We
furthermore assume a Salpeter IMF between 0.1 and 100 $M_{\odot}$  and a fixed
stellar metallicity of $Z_{\odot}/5$. For the old underlying population we
adopt an exponential SFR with an  e-folding time $\tau=3$ Gyr, except for
model \texttt{c2} for which $\tau$ was  set to 1 Gyr. For these two e-folding
times the observed $B$-$R$ color of the host galaxy is reproduced by a present
age of 10.3 Gyr and 5 Gyr,  respectively.  Models \texttt{inst} and \texttt{c0}
fit both the stellar SED and the observed  EWs of the \ha\ and \hb\ emission
lines whereas in models  \texttt{c1} and \texttt{c2} only the stellar SED was
fitted.

The results for each SF region are summarized in Table \ref{tab:evol}. The age
and mass of the young stellar component $t_{\rm SF}$ and  $M_{\rm \star,SF}$
are given in cols. 3 and 4, respectively.  Columns 5 and 6 list the predicted
EWs of the \Ha\ and \Hb\ emission line and $B-R$ color. Column 7
gives the difference $\Delta\mu_B$ between the $B$-band surface brightness
of the old stellar population and  that of the total stellar population
predicted by the fit ($\mu_{B,{\rm old}}-\mu_{B,{\rm total}}$), and col. 8 the
reduced $\chi^2$. The observed EWs of the \Ha\ and \Hb\ lines
are also indicated for each SF region.

The best-fitting stellar SEDs for each circumnuclear region, overlaid with the
observed spectrum, are shown in Fig.~\ref{Fig:Starlight} (bottom panel). The
SEDs corresponding to models \texttt{inst}, \texttt{c0} and \texttt{c1} are
shown with the black, solid-gray and dashed-gray curve, respectively.  The
blue and red curves illustrate the young and old stellar component,  as
derived from the \texttt{inst} model.

Inspection of Table \ref{tab:evol} and Fig.~\ref{Fig:Starlight}
shows that the instantaneous model reproduces best both the stellar SED and 
the EW of \Ha\ and \Hb, implying in all cases a burst age 
$t_{\rm SF}$ of 6 to 10 Myr. 
The predicted stellar mass from this model adds up to $\sim 1.5\times 10^6$.
Models \texttt{c0}, assuming continuous star formation since t$_{\rm SF}$ 
(in most cases $\geq 2$ Gyr), generally fail to simultaneously reproduce both 
the EWs and the stellar SED, with a clear tendency for underestimating the 
latter for $\lambda<4500$ \AA. 
Continuous SF models fitting the stellar SED only (\texttt{c1} and 
\texttt{c2}) predict, similar to the instantaneous model, a young age, 
however, they strongly overestimate the EWs of of \Ha\ and \Hb.
Comparison of models \texttt{c1} and \texttt{c2} shows that the adopted
e-folding time for the host galaxy has a minor effect on t$_{\rm SF}$.

\begin{deluxetable}{lcccccccc}
\tabletypesize{\footnotesize}
\tablecaption{Model results for each SF region}
\tablehead{
\colhead{Region}           & \colhead{Model}               & 
\colhead{t$_\mathrm{SF}$}  & \colhead{$M_\mathrm{*,SF}$}   & 
\colhead{EW(\Ha)}          & \colhead{EW(\Hb)}             & 
\colhead{$B-R$}            & \colhead{$\Delta\mu_B$}       & 
\colhead{$\chi^2$} 
\\
\colhead{}                 & \colhead{}                    &
\colhead{}		   & \colhead{($10^6$ M$_{\sun})$} & 
\colhead{\AA}              & \colhead{\AA}                 & 
\colhead{(mag)}            & \colhead{(mag)}               & 
\colhead{} 
}
\startdata
1   & inst &    10\phd\phn\phn Myr & 0.27    & \phn20\phd\phn     & \phn5\phd\phn   & 0.82    & 0.62 & 0.044 \\
    & c0   & \phn2.5\phn       Gyr & 1.46    & \phn23\phd\phn     & \phn7\phd\phn   & 1.05    & 0.17 & 0.389 \\
    & c1   & \phn7\phd\phn\phn Myr & 0.11    &    296\phd\phn     &    78\phd\phn   & 0.84    & 0.74 & 0.029 \\ 
    & c2   &    10\phd\phn\phn Myr & 0.09    &    237\phd\phn     &    61\phd\phn   & 0.82    & 0.75 & 0.032 \\
    & obs  &	                   &         & \phn19.7           & \phn5.4         &         &      &       \\
\hline
%
2   & inst & \phn8\phd\phn\phn Myr & 0.17    & \phn28\phd\phn     & \phn7\phd\phn   & 0.80    & 0.60 & 0.047 \\
    & c0   & \phn2.5\phn       Gyr & 2.8\phn & \phn34\phd\phn     &    10\phd\phn   & 0.96    & 0.42 & 0.218 \\
    & c1   & \phn7\phd\phn\phn Myr & 0.11    &    292\phd\phn     &    77\phd\phn   & 0.84    & 0.73 & 0.024 \\
    & c2   &    10\phd\phn\phn Myr & 0.09    &    235\phd\phn     &    60\phd\phn   & 0.82    & 0.74 & 0.025 \\
    & obs  &	                   &         & \phn28.9           & \phn7.9         &         &      &       \\
\hline
3   & inst &    44\phd\phn\phn Myr & 37.4    & \phn\phn7\phd\phn  & \phn2\phd\phn   & 0.60    & 1.48    & 0.227     \\
    & c0   & \nodata               & \nodata & \nodata            & \nodata         & \nodata & \nodata &  \nodata  \\
    & c1   & \phn2\phd\phn\phn Gyr & 49.1    &       108\phd\phn  & 24\phd\phn      & 0.56    & 2.9     & 0.121     \\
    & c2   & \phn1\phd\phn\phn Gyr & 47.3    &       112\phd\phn  & 25\phd\phn      & 0.57    & 1.9     & 0.125     \\
    & obs  &	                   & 	     & \phn\phn7.5        & \phn1.4         &         &         &           \\
\hline
4   & inst & \phn8\phd\phn\phn Myr & 0.27    & \phn33\phd\phn     & \phn8\phd\phn   & 0.70    & 0.80 & 0.085 \\
    & c0   & \phn2.5\phn       Gyr & 4.8\phn & \phn44\phd\phn     &    12\phd\phn   & 0.89    & 0.64 & 0.374 \\
    & c1   &    10\phd\phn\phn Myr & 0.19    &    326\phd\phn     &    80\phd\phn   & 0.75    & 0.97 & 0.031 \\
    & c2   &    20\phd\phn\phn Myr & 0.20    &    248\phd\phn     &    60\phd\phn   & 0.73    & 0.97 & 0.033 \\
    & obs  &	                   &         & \phn35.6           & \phn9.1         &         &      &       \\
\hline
5   & inst & \phn8\phd\phn\phn Myr & 0.28    & \phn31\phd\phn     & \phn8\phd\phn   & 0.73    & 0.74 & 0.053 \\
    & c0   & \phn3\phd\phn\phn Gyr & 4.4\phn & \phn36\phd\phn     &    10\phd\phn   & 0.95    & 0.47 & 0.380 \\
    & c1   &    30\phd\phn\phn Myr & 0.38    &    197\phd\phn     &    49\phd\phn   & 0.76    & 0.86 & 0.025 \\
    & c2   &    17\phd\phn\phn Myr & 0.18    &    241\phd\phn     &    60\phd\phn   & 0.76    & 0.89 & 0.024 \\
    & obs  &	               &             & \phn29.9           & \phn7.4         &         &      &       \\
\hline
6   & inst & \phn6\phd\phn\phn Myr & 0.47    & \phn55\phd\phn	  & 14\phd\phn	    & 0.76    & 0.62 & 0.092 \\
    & c0   & \phn0.45	       Gyr & 3.7\phn & \phn65\phd\phn	  & 17\phd\phn	    & 0.84    & 0.63 & 0.067 \\
    & c1   & \phn7\phd\phn\phn Myr & 0.34    &    279\phd\phn	  & 74\phd\phn	    & 0.85    & 0.70 & 0.027 \\
    & c2   & \phn7\phd\phn\phn Myr & 0.24    &    280\phd\phn	  & 73\phd\phn	    & 0.84    & 0.72 & 0.029 \\
    & obs  &                       &         & \phn60.1           & 18.6            &         &      &       \\
\hline
\enddata
\tablecomments{Col. 1: Star forming region ID; col. 2: model; cols. 3, 4: age and
mass of the young stellar component; cols 5, 6: predicted equivalent width of
the \Ha\ and \Hb\ emission lines; col. 7: difference between the $B$-band
surface brightness of the older stellar population and that of the total
stellar population predicted by the fit; col. 8: reduced $\chi^2$ of the fit.
For each knot, the last line, labeled ``obs'', shows the observed \Ha\ and
\Hb\ equivalent widths.} 
\label{tab:evol}
\end{deluxetable}

In summary, both population and evolutionary spectral synthesis models argue
independently against the idea that the SF ring in Mrk~409 is forming
continuously with a constant SFR since several $10^8$ yr, as expected  from the
RR scenario. Instead, they consistently point to a recent  ($\sim$10 Myr)
enhancement of the SFR in all circumnuclear SF regions. 

Clearly, regardless of the model used, any conclusions related  to the SFH of
the ring should be considered with caution. It is known that both spectral
synthesis approaches are plagued by ambiguities, especially when applied to
complex emitting sources, such as starburst galaxies, involving multiple 
stellar populations of different age and metallicity, and being likely 
subject to different amounts of intrinsic extinction.

Solutions by {\sc starlight} are guided by the minimal linear combination  of
age vectors, i.e. they discard out of the available library  SSPs with a low
significance. Aside from the degeneracy problem,  the solutions may therefore
provide a very fragmentary view of the true SFH  in a galaxy. In our concrete
case, the absence of SSPs in the age interval  between $\sim 40$ Myr and $\sim
1$ Gyr should not be taken as foolproof of cessation of SF activities during
that period, and the hypothesis of a low-level star formation in the ring
cannot be rejected with certainty.

Evolutionary synthesis models suffer as well from ambiguities stemming 
not only from the age-metallicity degeneracy but, as thoroughly discussed
in \cite{Guseva2001}, by uncertainties in the intrinsic extinction
and in the SFH assumed. The instantaneous and continuous model likely
represent two extreme cases and rather poor approximations to the 
true SFH along the circumnuclear ring.
For example, a continuous SF process over the last $\sim 1$ Gyr 
with a varying or recently elevated SFR may also match the observations.
However, an investigation of more complex SFHs on the basis of the
present data would likely not completely lift uncertainties 
in the interpretation of the SF ring.

We stress that the satisfactorily agreement between the two conceptually
different models used is reassuring. We are therefore  confident in that an
instantaneous burst is a better description of the SFH of the individual SF
knots in the ring. 

Furthermore, the morphology of the galaxy does not show any bar or
non-axisymmetric features, conditions required for the formation of a galactic
RR. In a recent paper, \cite{Comeron2009} note that Mrk~409 (NGC~3011) has no
obvious bar, and place it into the group of galaxies that have ``nuclear rings
that cannot easily be explained in a resonance framework''.  

Hence, both the synthesis models results and the galaxy morphology speak against
the presence of a RR in this galaxy, and  favor the starburst-driven shock
scenario as the most plausible to explain the origin of the SF ring in Mrk~409.

\begin{figure}   
\centerline{  
\hspace*{0.0cm}\includegraphics[angle=-90.0,viewport=180 40 520 550,
clip,width=1.00\textwidth]{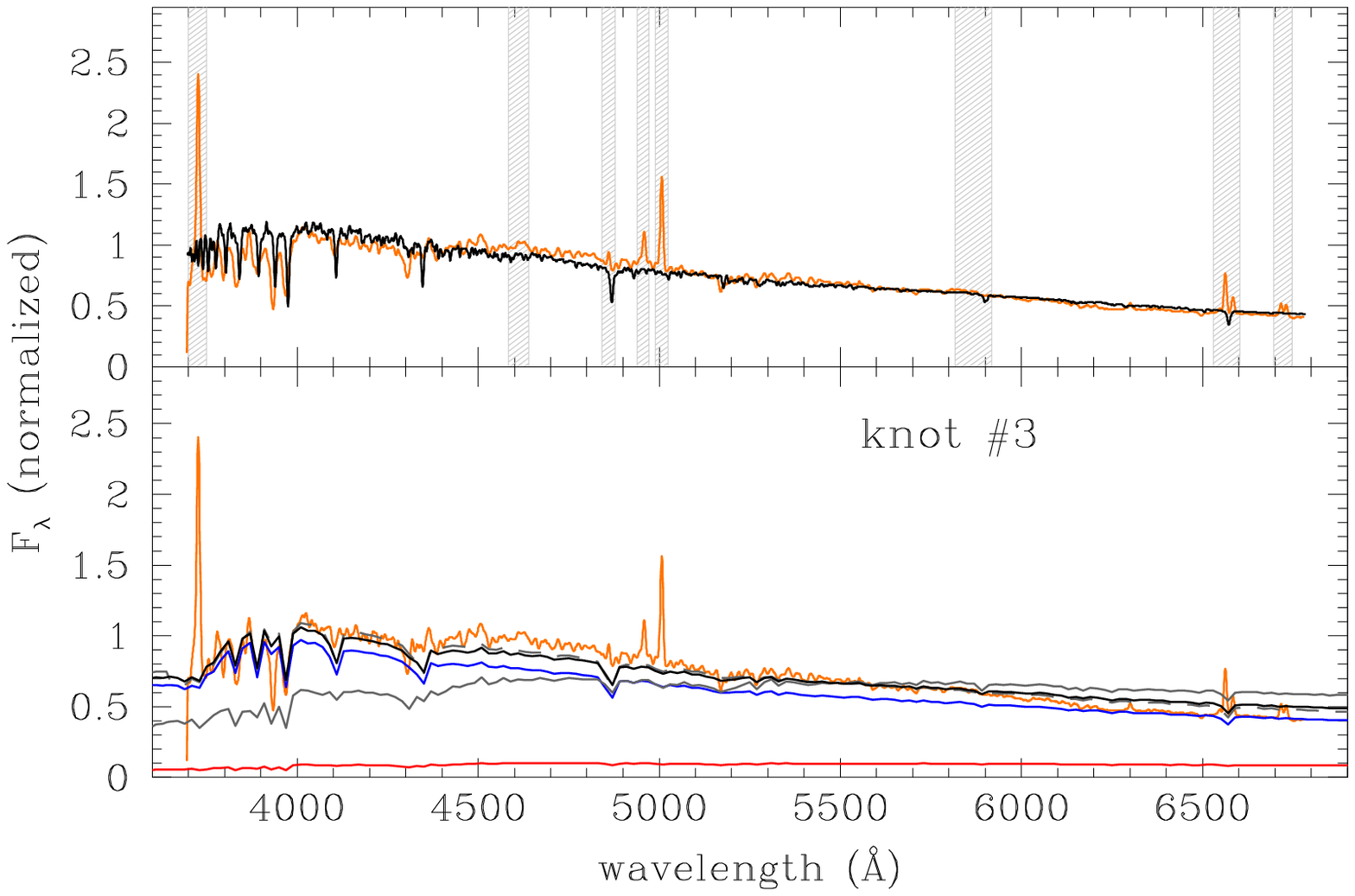}
}
\caption{Observed and best-fitting synthetic spectrum with STARLIGHT (upper
panel) and Pegase~2.0 (bottom panel) of the nuclear SF region \#3. It can be
seen from either diagram that a stellar population cannot fit the observed
stellar continuum without systematic residuals. }
\label{Fig:model_region3}
\end{figure}

The nature of the nuclear region \#3 is puzzling. 
As discussed already, its observed line ratios suggest an additional
non-thermal source. 
Further support to this interpretation comes from spectral synthesis models 
which both fail to reproduce its observed SED without large systematic
residuals (cf. Fig.~\ref{Fig:model_region3}). 
Although inclusion of a power law component along with a variation in the 
extinction model might produce a better fit, we preferred not to do so, as 
this would probably not add solid constraints to the understanding of the 
nature and SFH of the nuclear region.

However, assuming that the luminosity of region \#3 is dominated by stars, we
can formally check whether its  energetic output is consistent with the blast
wave scenario. Taking its predicted burst age and total stellar mass of 
$t_{\rm SF}\approx 44$ Myr and $M_{\star,{\rm SF}}\approx 3.7 \times 10^7$
M$_{\sun}$  (Table \ref{tab:evol}) at face value, we estimate from {\sc
starburst99} \citep[][their Fig. 111]{Leitherer1999} the mean
\emph{mechanical} luminosity  imparted to the ISM over $t_{\rm SF}$ via
radiative winds and SNe to $\sim3.8\times 10^{41}$ erg s$^{-1}$. This modest
mechanical luminosity, which is to be taken as an upper limit, can indeed
readily produce a super-shell of the observed dimensions within a few Myrs.

From Eq.~2 of \cite{McCrayKafatos1987}, the radius in pc of a 
shell writes as  $R_{\mathrm s}=269 (L_{38}/n_0)^{1/5} \cdot t_7^{3/5}$, 
where $L_{38}$  denotes the mechanical luminosity in units of $10^{38}$
erg/s, $t_7=t/(10^7\,{\mathrm yr})$ and $n_0$ is the density of the ambient
gas in cm$^{-3}$. 
For $L_{38} \sim 3800$ the radius of the starburst-driven shell 
at the age $t_{\rm SF}$ is $R_{\mathrm s}\approx$3400/$n_0^{1/5}$.
Hence, the energetic output from the nuclear starburst region is sufficient 
for producing within $t\leq t_{\rm SF}$ a super-shell with a $R_\mathrm{s}$ 
between  $\sim$600 pc (inner SF ring) and $\sim$2.3 kpc (outer SF ring),
depending on $n_0$.
 
While the simple blast-wave scenario can formally be reconciled with the 
known properties of Mrk~409, a major open question pertains to the actual 
SFH of region \#3.

While a conclusive assessment of this issue is out of reach with the present
data, the co-evolution of a massive nuclear SF region and an AGN remains a
viable and appealing interpretation for the overall characteristics of Mrk~409.

\section{Conclusions}
\label{Sect:Conclusions}


We presented PMAS integral field spectroscopy of the central
$16\arcsec\times16\arcsec$ of the BCD galaxy Mrk~409.  Mrk~409 is known to
undergo intense SF  activity within the central part of an old,  elliptical host
galaxy \citep{GildePaz2005}. Previous work (GdP03) had revealed a compact
(diameter $<5\arcsec$) nuclear starburst region, surrounded by two nearly
concentric rings of ionized gas emission with radii of 0.6 kpc and 2.8 kpc. The
inner ring was proposed to be the result of in situ star formation in a dense
shell of piled-up material on the surface of an expanding starburst-driven
super-bubble. 

The main results from the present study may be summarized as follows:   

\begin{itemize}

\item The inner circumnuclear ring of Mrk~409 contains a young stellar
population with a total mass of $\sim 1.5\times 10^6$ \msun\ , which 
likely started forming almost coevally $\sim 10^7$ yr ago. This is
strong evidence to the interpretation that circumnuclear SF activities in
Mrk~409 were triggered through the collision of a spherically expanding,
starburst-driven super-bubble with the ambient denser material. Although the
details of this process remain unclear, our study suggests that sequential
star formation in a spherical shell, as observed in Mrk~409, may contribute
considerably to the build-up of the young stellar component in BCDs. 

\item While the circumnuclear SF ring presents properties typical of 
\ion{H}{2} regions, diagnostic line ratios for the high-surface brightness
starburst nucleus suggest an additional ionization source for this region,
most probably a low-luminosity AGN.

\item The nuclear and circumnuclear SF components in Mrk~409 also differ
considerably in their extinction properties: SF regions in the ring show a
moderate and relatively uniform intrinsic extinction ($0.2\la\chbeta\la0.3$),
while the nuclear emission is subject to strong dust obscuration
($\chbeta\approx0.9$). Consequently, corrections for extinction based on the
luminosity-weighted spectrum of the central region of a BCD can lead, in the
case of Mrk~409, to a severe over-estimation of the \Ha\ luminosity and star
formation rate. 

\item  Mrk~409 shows a relatively high oxygen abundance
$(12+\log(\mathrm{O/H})\sim8.40$), without evidence for any significant 
chemical abundance gradient out to $R\sim 0.6$ kpc. 

\item All the studied circumnuclear \ion{H}{2} regions have electron
densities in the low density limit. In the nuclear starburst region the
electron density increases to a value as large as 500 cm$^{-3}$.

\item The ionized gas component within the inner $\sim 1$ kpc of Mrk~409
displays a smooth kinematical pattern which is likely dominated by rotation;
the total mass inside the SF ring is estimated to be 
$\simeq 1.4 \times 10^9$ $M_\sun$.

\end{itemize}

\acknowledgments

LMC and CK acknowledge the Alexander von Humboldt Foundation. NC and CZ are
grateful for the hospitality of the Astrophysikalisches Institut Potsdam. We
thank Bego{\~n}a Garc{\'\i}a-Lorenzo and Ana Monreal-Ibero for useful
suggestions. We also thank Jos{\'e}~N. Gonz{\'a}lez-P{\'e}rez, Jos{\'e}~M.
V{\'\i}lchez and Inma Mart{\'\i}nez-Valpuesta for fruitful discussions. This
research has made use of the NASA/IPAC Extragalactic Database (NED), which is
operated by the Jet Propulsion Laboratory, Caltech, under contract with the
National Aeronautics and Space Administration. The authors acknowledge the
work of the Sloan Digital Sky Survey (SDSS) team. Funding for the SDSS has
been provided by the Alfred P. Sloan Foundation, the Participating
Institutions, the National Aeronautics and Space Administration, the National
Science Foundation, the U.S. Department of Energy, the Japanese
Monbukagakusho, and the Max Planck Society. The SDSS Web site is
http://www.sdss.org/. This work has been partially funded by the Spanish
``Ministerio de Ciencia e Innovaci{\'o}n'' through grants AYA 2007 67965 and
HA2006-0032, and under the Consolider-Ingenio 2010 Program grant
CSD2006-00070: First Science with the GTC
(http://www.iac.es/consolider-ingenio-gtc/).

\end{document}